\documentstyle[aps,twocolumn,epsf,amssymb]{revtex}
\def\ave#1{\langle #1\rangle}
\def\C{\Bbb{C}}
\newcommand{\ma}[1]{{\rm\bf #1}} 
\newcommand{\ve}[1]{{\vec #1}}

\newcommand{\bra}[1]{\langle #1|}
\newcommand{\ket}[1]{|#1\rangle}
\newcommand{\braket}[2]{\langle #1|#2\rangle}
\newcommand{\half}{{\textstyle{\frac{1}{2}}}}
\newcommand{\quar}{{\textstyle{\frac{1}{4}}}}
\newcommand{\oneL}{{\textstyle{\frac{1}{L}}}}
\newcommand{\ad}{{\rm ad\,}}
\newcommand{\tr}{{\rm tr\,}}
\begin{document}
\draft
\twocolumn[\hsize\textwidth\columnwidth\hsize\csname %
@twocolumnfalse\endcsname
\title{Ergodic properties of a generic non-integrable
quantum many-body system in thermodynamic limit}
\author{Toma\v z Prosen}
\address{Physics Department, Faculty of Mathematics and Physics,
University of Ljubljana, Jadranska 19, 1111 Ljubljana, Slovenia
}  
\date{\today}
\draft
\maketitle
\begin{abstract}
\widetext
We study a generic but simple non-integrable quantum 
{\em many-body} system of {\em locally} interacting particles, namely a 
kicked $t-V$ model of spinless fermions on 1-dim lattice 
(equivalent to a kicked Heisenberg XX-Z chain of $1/2$ spins). 
Statistical properties of dynamics (quantum ergodicity and quantum mixing) 
and the nature of quantum transport in {\em thermodynamic limit} are considered 
as the kick parameters (which control the degree of non-integrability) are varied.
We find and demonstrate {\em ballistic} transport 
and non-ergodic, non-mixing dynamics (implying infinite conductivity at 
all temperatures) in the {\em integrable} regime of zero or very small kick 
parameters, and more generally and important, also in {\em non-integrable} regime 
of {\em intermediate} values of kicked parameters, 
whereas only for sufficiently large kick parameters we recover quantum ergodicity 
and mixing implying normal (diffusive) transport. 
We propose an order parameter (charge stiffness $D$) which controls the phase
transition from non-mixing/non-ergodic dynamics (ordered phase, $D>0$) to
mixing/ergodic dynamics (disordered phase, $D=0$) in the thermodynamic limit.
Furthermore, we find {\em exponential decay of time-correlation function} in
the regime of mixing dynamics.

The results are obtained consistently within three different numerical and 
analytical approaches: (i) time evolution of a finite system and direct computation 
of time correlation functions, (ii) full diagonalization 
of finite systems and statistical analysis of stationary data, 
and (iii) algebraic construction of quantum invariants of motion of an infinite system, 
in particular the time averaged observables.
\end{abstract}

\bigskip
]
\narrowtext

\section{Introduction}

It has been a common belief for a long time, that a large
system of sufficiently many interacting particles should fill uniformly 
the entire available phase space.
This is known as {\em ergodic hypothesis}, one of the cornerstones
of the statistical mechanics, and is a necessary assumption to justify
the use of canonical ensembles and derivation of fundamental laws of
statistical physics, such as transport laws (e.g. Ohm's law or
Fourier's law).

However, the proof together with the precise conditions for the validity of ergodic 
hypothesis is still one of the most fundamental unsolved problems of theoretical 
physics. Even in the context of purely classical dynamics, the ergodic theory
\cite{Sinai,Arnold68},
though it is an involved and beautiful mathematical discipline, 
can make strong statements only for a very limited class of systems, while generic 
dynamical systems, especially those consisting of many interacting 
particles, are far from being understood \cite{Henon,FPU}.
Even less is known about ergodic properties of generic {\em quantum} 
many body systems, which is precisely the objective of this paper.
A closed (finite) and bounded quantum system of size $L$ and with a finite number 
$N$ of particles has a discrete spectrum, 
hence its time evolution is quasi-periodic, and
accordingly it is non-ergodic and non-mixing, as we shall define below.
However, in the thermodynamic limit (TL), of diverging
size $L\rightarrow \infty$ and density 
of particles $\rho=N/L$ fixed, the spectrum of the quantum propagator may accuire
a continuous component, and one may expect genuine properties of quantum ergodicity 
and quantum mixing to set in provided the strength of non-linear interaction
is sufficiently strong.
In this paper we deal with general non-autonomous many-body systems with
Hamiltonians $H(\tau)$ which explicitly depend on time $\tau$.
Therefore the entire Hilbert space of many-body quantum configurations 
(Fock space) is dynamically accessible and the `micro-canonical' average of an 
{\em intensive} observable represented by an operator $A$, reads
\begin{equation}
\ave{A} = \lim_{L\rightarrow\infty} \frac{\tr A}{\tr 1}.
\label{eq:mc1}
\end{equation}
If the system possesses a group of exact geometric or dynamical symmetries,
the trace in eq. (\ref{eq:mc1}) may be considered only over a specific
symmetry class of the Fock space w.r.t. symmetry group. For example,
if the system is autonomous, energy is conserved, and eq. (\ref{eq:mc1}) should 
be replaced by the average over
a specific ``energy shell'', or, as often, if the number $N$ of particles
(or the particle density $\rho=N/L$) is preserved, then the micro-canonical 
average should be performed over the Fock supscace of N-particle configurations
\begin{equation}
\ave{A}_\rho = \lim_{L\rightarrow\infty} \frac{\tr(A\delta_{[\rho L],N})}
{\tr \delta_{[\rho L],N}}
\label{eq:mc2}
\end{equation}
where $[x]$ is an integer part of $x$ and $\delta_{m,n}$ is a Kronecker symbol.
When we will like to keep the size $L$ in the average (\ref{eq:mc2}) fixed and
finite we
would write $\ave{A}^L_\rho$. Although in this abstract discussion we would like 
to avoid the notion of temperature \cite{Temp}, one may think of (\ref{eq:mc1}) 
or (\ref{eq:mc2}) as canonical averages at very large or infinite temperature,
$\beta = (k_B T)^{-1}\rightarrow 0$.

The system is {\em quantum ergodic} if the time average of an arbitrary observable
in Heisenberg picture $A(\tau)$ equals to the micro-canonical average
$\ave{A}$ times a unit operator (over the corresponding 
desymmetrized Fock subspace)
\begin{equation}
\bar{A}:=\lim_{\rightarrow\infty}\frac{1}{T}\int_0^T d\tau A(\tau) = \ave{A} 1.
\label{eq:erg1}
\end{equation}
In case where one has a constant of motion, e.g. density $\rho$ (or energy $E=H$, etc.)
one should define the ergodicity through the spectral resolution of
the relevant invariant operator, $\rho = \int \rho' d E_{\rho'}$, namely
\begin{equation}
\bar{A} = \int \ave{A}_{\rho'} d E_{\rho'}.
\label{eq:erg2}
\end{equation}
When the micro-canonical average does not depend on the 
eigenvalues of the symmetry operations or {\em quantum numbers},
e.g. $\ave{A}_\rho = \ave{A}$, definition (\ref{eq:erg2}) is equivalent to a 
simple one (\ref{eq:erg1}).  

Even stronger ergodic property is quantum mixing, which is defined very generally
\cite{JLP} as follows: The infinite ($L=\infty$) quantum many-body system 
is called quantum mixing if
time-correlations for an arbitrary pair of quantum observables in Heisenberg
representation, $A(\tau)$ and $B(\tau)$, decay to zero
\begin{eqnarray}
C_{AB}(\tau) &:=& \ave{A(\tau)B(0)} - \ave{A}\ave{B}, \nonumber \\
\lim_{\tau\rightarrow\infty} C_{AB}(\tau) &=& 0.
\label{eq:mix}
\end{eqnarray}
Note that formula (\ref{eq:mix}) implies that TL should be considered prior
to the time limit, $\tau\rightarrow\infty$, since these two limits do not 
generally commute \cite{JLP}.
Again, in case of additional symmetry, mixing over separate 
symmetry classes can be studied; as well as {\em uniform mixing}
over entire Fock space (which makes sense when canonical averages
of $A$ and $B$ do not depend on quantum numbers like $\rho$).

In Ref. \cite{JLP} quantum mixing of a system of interacting bosons
has been related to a hard chaos of the corresponding classical 
(mean field) model.
However, general quantum systems need not possess the classical limit; 
when they do, the general definitions (\ref{eq:erg1}-\ref{eq:mix})
go over to the correct definitions of classical ergodicity and mixing  
of the corresponding classical counterparts.

It is easy to see that the same implication holds as in
classical mechanics \cite{Arnold68}: 
Quantum mixing (\ref{eq:mix}) implies quantum ergodicity (\ref{eq:erg1}).
To see this, observe that as a simple consequence of (\ref{eq:mix}),
time averaged correlation function should vanish,
$$\bar{C}_{AB} := \lim_{T\rightarrow\infty}\frac{1}{T}\int_0^T d\tau C_{AB}(\tau) = 0.$$
This fact is equivalent to
$$ \ave{\bar{A} B} - \ave{A}\ave{B} = \ave{(\bar{A} - \ave{A} 1)B} = 0.$$
Then we immediately see that observable in brackets should vanish
$\bar{A} - \ave{A} 1 = 0$, since observable $B$ is arbitrary. q.e.d.
The last argument can be reversed, so we see that 
quantum ergodicity (\ref{eq:erg1}) is equivalent to
$\bar{C}_{AB}=0$ for an arbitrary pair $A,B$.

We expect that quantum mixing implies universal statistical properties
of energy spectra (and also universal statistics of occupation numbers
\cite{Flambaum}, matrix elements, etc) described by 
Random Matrix Theory\cite{Mehta}, (RMT). Random matrix spectral statistics 
have indeed been demonstrated numerically for few strongly non-integrable 
many-body systems \cite{fRMT1}.
On the other hand, completely integrable quantum many-body systems
(having an infinite set of independent conservation laws $Q_n,n=1,2,3\ldots$)
are obviously non-ergodic ($\bar{Q_n} = Q_n \neq \ave{Q}_n 1$), 
and therefore non-mixing, and characterized by the universal Poissonian 
spectral statistics \cite{fRMT1}.

It has been pointed out recently \cite{Prelovsek,Zotos} that integrability typically 
implies non-vanishing stiffness, i.e. ideal conductance with infinite transport 
coefficients (or ideal insulating state).
Indeed, there is a direct implication of quantum mixing and quantum ergodicity
on quantum transport. One should simply inspect a Kubo formula \cite{Kubo},
which relates the real part of the transport coefficient, e.g. electric
conductivity $\sigma'(\omega)$, to the cosine transform of the
autocorrelation function of the electric current observable $J$,
written for high temperatures (small $\beta$) as
\begin{equation}
\sigma'(\omega) = 
\half\beta \int_{-\infty}^\infty \cos(\omega\tau) 
\ave{\oneL J(0)J(\tau)} d\tau.
\label{eq:kubo}
\end{equation}
The transport is diffusive and the system behaves as a {\em normal
conductor} if zero-frequency (d.c.) conductivity is finite,
$\sigma'(0) < \infty$, which means that the time integral of the
current-current correlation function should be finite; this is true
if the system is mixing and if time correlations decay sufficiently
fast, e.g. it is sufficient that 
$|\ave{\frac{1}{L}J(0)J(\tau)}| < C |\tau|^{-\alpha}$ 
for some $C > 0$ and $\alpha > 1$.
 
On the other hand, if the transport is ballistic, d.c. conductivity
diverges $\sigma'(0) = \infty$ and frequency dependent conductivity
can be written as a sum of delta-spike and a regularized 
conductivity 
\begin{equation}
\sigma'(\omega) = D \delta(\omega) + \sigma_{\rm reg}(\omega),
\quad 
D = \lim_{\epsilon\rightarrow 0}\int_{-\epsilon}^\epsilon \sigma'(\omega)d\omega
\end{equation}
The weight $D$ is known as a {\em charge stiffness} (or Drude weight)
and is proportional to the averaged current-current time correlator
\begin{equation}
D = \beta D_J, \quad
D_A = \lim_{T\rightarrow\infty}\lim_{L\rightarrow\infty} 
\frac{1}{2 T L} \int_{-T}^T C^L_{AA}(\tau) d\tau
\label{eq:stif}
\end{equation}
Therefore, non-vanishing charge stiffness $D_J \neq 0$, meaning a ballistic
electric transport, is a sufficient condition for deviation from quantum 
ergodicity, so $D_J$ will be extensively used as a quantitative indicator of 
quantum (non)ergodicity throughout the rest of this paper.

In a generic integrable system one can find an (infinite) set of
invariant {\em extensive} observables, the so-called conserved charges $Q_n$. 
Mazur \cite{Mazur} and Suzuki \cite{Suzuki} have proposed
a `Parseval-like' inequality for the time averaged autocorrelator of any 
extensive observable $A$
\begin{equation}
D_A \ge \sum_n \frac{|\ave{\oneL A Q_n}|^2}{\ave{\oneL Q_n^2}},
\label{eq:Mazur}
\end{equation}
using any suitable (sub)set of conserved charges $\{Q_m\}$, such that
$\ave{\oneL Q_n Q_m} = 0$ if $n \neq m$.
For an integrable system one therefore proves ideal (ballistic) transport
($D_J > 0$) if at least one term in an infinite sum on RHS of eq. (\ref{eq:Mazur}) 
applied to the current observable $A=J$ is non-vanishing (what is typically the 
case) \cite{Zotos}.
It is convenient to say that $\{ Q_n\}$ is a {\em complete} set of conserved
charges if (\ref{eq:Mazur}) is an exact {\em equality} for any observable 
$A$ (which is `square-summable', $\ave{\oneL A^2} < \infty$).

The important and delicate question is whether non-ergodicity
of an integrable system in TL can be structurally stable against 
generic and finite non-integrable perturbation.
In this paper we will present clear numerical evidences
based on various different and independent numerical methods in support 
of a conjecture claiming an affirmative answer to the above question.
\\\\
{\bf Conjecture:} Let $H_\lambda$ be a continuous family of 
{\em generic infinite} quantum-many body systems with {\em local}
interaction \cite{Conj}, such that $H_0$ is completely-integrable while
$H_\lambda$ are non-integrable for almost any $\lambda\neq 0$. 
Then, $\exists \lambda_c$, such that $H_\lambda$ are {\em non-ergodic} and 
{\em non-mixing} for $|\lambda| < \lambda_c$.
\\\\
One important consequence of the above Conjecture would be 
a large class of not only completely integrable but also nearly-integrable
many-body systems for which ideal transport and infinite conductance
would be expected.

In Sec. II we define a two parametric family of generic non-integrable 
many-body systems \cite{Prosen1}, 
namely a kicked t-V model of interacting spinless fermions,
or equivalently, a kicked Heisenberg XX-Z spin-$\half$ chain,
and describe the basic properties of the model.
In Sec. III we show how efficient explicit time evolution
of the above model of finite but quite large size $L$ can be computed, 
and present results on extensive numerical computation of time 
correlation functions. By letting the size 
$L$ to increase and inspecting TL, we 
clearly identify two regimes of quantum motion: 
{\em non-mixing} regime for small and intermediate
values of kick parameters where time-correlation functions typically saturate to
constant non-vanishing values, and {\em exponentially mixing} regime for 
sufficiently large values of kick parameters where time correlation functions decay 
{\em exponentially}. As a complementary approach to direct time evolution 
(time domain) we perform, in Sec.IV, complete diagonalization of stationary 
problem (frequency domain) for finite sizes $L$. Using stationary data we have 
computed and analyzed short-range and long-range quasi-energy level statistics. 
By means of matrix elements of the current observable $J$ we have also directly 
computed the conductance and the charge stiffness $D_J$. The obtained results
are in quantitative agreement with a direct time evolution (Sec.III).
In Sec. V we outline completely different and independent method of
computing time-averaged observables and quantitative indicators of
quantum ergodicity such as the charge stiffness $D_J$, by making 
use of extensive computerized Lie algebra.
This third method, in contrast to other two, refers directly to infinite
systems (infinite lattices $L=\infty$) and, again, gives compatible 
results. In Sec. VI we give more arguments in support of our Conjecture by 
discussing relevant published and unpublished results and conclude.

\section{The Model}

No general analytical methods exist to deal with dynamics
of non-integrable quantum many-body systems. A beautiful solition theory
based on inverse scattering and algebraic Bethe ansatz\cite{KorepinBook} 
is unfortunately applicable only to a limited class of very special,
namely completely integrable many-body systems.
Therefore one is left to numerical experiment to learn about 
dynamics of generic quantum systems, encouraged with the fact that 
numerical and experimental investigations of dynamical systems of one or
few particles has been a very fruitful area of research (known as Quantum Chaos)
over the past twenty years \cite{QChaos}.
However, one should be very careful in picking out the toy model, since the
fact, that the dimensionally of the Hilbert space (Fock space of quantum 
many-body states) grows exponentially with increasing system size $L$, makes 
a serious quantitative study of TL almost prohibitive.
Here we propose the simplest many-body system that we can think of:
one dimensional lattice of spinless fermions of size $L$, 
and for the reasons which will become clear in the next section, 
we decide to break integrability by taking time-dependent 
interaction which is switched on periodically by means of $\delta$-kicks.
Therefore, the time-dependent Hamiltonian of our ``kicked t-V model''
(KtV)\cite{Prosen1} reads
\begin{equation}
H(\tau) = \sum_{j=0}^{L-1} \left[-\half t(c^\dagger_j c_{j+1} + h.c.) +
\delta_p(\tau) V n_j n_{j+1}\right].
\label{eq:KtV1}
\end{equation}
$c^\dagger_j$, $c_j$ are fermionic creation and
annihilation operators satisfying canonical anticommutation
relations $[c_j,c_k]_+:=c_j c_k + c_k c_j =0,
[c^\dagger_j,c_k]_+=\delta_{jk}$, $n_j=c^\dagger_j c_j$ are
number operators, and periodic boundary conditions are imposed
$c_L\equiv c_0$.
$\delta_p(\tau) = \sum_{m=-\infty}^\infty \delta(\tau-m)$ is
a periodic $\delta$-function. We use units in which $\hbar=
({\rm time\;between\;collisions})=({\rm lattice\;spacing})=1$.
The hopping amplitude $t$ and interaction strength
$V$ are independent (kick) parameters.
An important and useful property of kicked 
systems like (\ref{eq:KtV1}) is
the fact that the evolution propagator over one period (Floquet operator)
factorizes into the product of kinetic and potential part
\begin{eqnarray}
U &=& \hat{\cal T}\exp\left(-i\int_{0^+}^{1^+} d\tau H(\tau)\right) \nonumber\\
  &=& \exp(-i W)\exp(-i T), \label{eq:mbmap}\\
T &=& -{\half} t \sum\limits_{j=0}^{L-1} 
\left(e^{i\phi} c^\dagger_{j+1} c_j  
+ e^{-i\phi} c^\dagger_j c_{j+1}\right), \\
W &=& V \sum\limits_{j=0}^{L-1} n_j n_{j+1}. \label{eq:WW}
\end{eqnarray}
We have used a Pierels phase $\phi$ in order to introduce a 
particle current
\begin{equation}
J = (i/t) U^\dagger (\partial/\partial\phi) U\vert|_{\phi=0} =
\half i \sum\limits_{k=0}^{L-1}
\left(c^\dagger_j c_{j+1} - c^\dagger_{j+1} c_j\right)
\label{eq:cur}
\end{equation}
(which is divided by $t$ for convenience); elsewhere we put 
$\phi:=0$.
Note that the kinetic energy $T$ as well as
the current $J$ are {\em diagonal} in momentum representation
\begin{eqnarray}
T &=& t \sum_{k=0}^{L-1} (1-\cos(sk)) \tilde{n}_k, \label{eq:Tmom}\\
J &=&   \sum_{k=0}^{L-1} \sin(sk) \tilde{n}_k,     \label{eq:Jmom}		
\end{eqnarray}
where $s=2\pi/L$, while tilde refers to momentum representation of 
field operators
\begin{equation}
\tilde{c}_k = L^{-1/2} \sum_{j=0}^{L-1} \exp(i s j k) c_j,
\quad \tilde{n}_k = \tilde{c}^\dagger_k \tilde{c}_k.
\end{equation} 

Using a well know Jordan-Wigner transformation\cite{WignerJordan}
one can map 1-dim lattice of spinless fermions to a spin$-\half$ 
chain described by Pauli operators 
$\sigma^{\pm}_j=(\sigma^x_j \pm i \sigma^y_j)/\sqrt{2}$ and
$\sigma^z_j$, namely
\begin{eqnarray}
\sigma^+_j &=& \sqrt{2} 
c^\dagger_j \exp\left(i\pi \sum_{j'=0}^{j-1} n_{j'}\right) \nonumber \\
\sigma^z_j &=& 2 n_j -1 \label{eq:WigJor},
\end{eqnarray}
which satisfy canonical commutation relations 
$$
[\sigma^\mu_j,\sigma^\nu_k] = i\sum_\eta \epsilon_{\mu\nu\eta}\sigma^\eta_j
\delta_{jk},\quad
\mu,\nu,\eta\in\{x,y,z\}.
$$
In fact, Jordan-Wigner transformation (\ref{eq:WigJor}) maps KtV model on
a kicked Heisenberg XX-Z chain
\begin{eqnarray}
H(\tau) &=& T + \delta_p(\tau) W, \label{eq:XXZ} \\
T &=& \quar t \sum_{j=0}^{L-1} 
(\sigma^x_j \sigma^x_{j+1} + \sigma^y_j \sigma^y_{j+1}), \\
W &=& \quar V \sum_{j=0}^{L-1}
(\sigma^z_j \sigma^z_{j+1} + 2 \sigma^z_j).
\label{eq:KXXZW}
\end{eqnarray}
The last term of potential (\ref{eq:KXXZW}) is irrelevant 
since the total $z-$spin $S_z = \sum_{j=0}^{L-1} \sigma^z_j = 
N-\half L$ is a constant of motion, $[U,S_z]=0$.

Interaction strength $V$ is {\em cyclic parameter} 
$V\equiv V \pmod{2\pi}$, since
the spectrum of $W/V$ is a set of integers (see eq. (\ref{eq:WW})).  
KtV model is integrable/solvable in three special (limiting) cases:
\begin{enumerate}
\item $t=0$, 1-dim Ising model,
\item $V=0\pmod{2\pi}$, 1-dim free fermions, or equivalently, 1-dim Heisenberg XX 
$\half$-spin chain,
\item $tV\rightarrow 0$ and $\Delta=t/V$ finite, 
(continuous-time) 1-dim Heisenberg XXZ $\half$-spin chain.
\end{enumerate}
For $t\neq 0,V\neq 0 \pmod{2\pi}$, KtV is expected to be non-integrable,
possibly quantum ergodic and mixing. 

To conclude this section, let us list the symmetries of
a general KtV model (for arbitrary $t$ and $V$):
In addition to the trivial conservation law, namely the number or
density of particles 
\begin{equation}
N = \sum_{j=0}^{L-1} n_j,\quad \rho = \frac{N}{L},\quad [U,N]=0,
\end{equation}
and the total quasi-momentum $K\in\{0,1,\ldots,L-1\}$ which is defined  
as an eigenvalue of a unitary translational symmetry operation $S$
\begin{equation}
S = \exp(i s K) = 
\exp\left(i s \sum_{k=0}^{L-1} k \tilde{n}_k\right),\quad [U,S]=0,
\label{eq:transs}
\end{equation}
the KtV model has two (geometric) `reflection' symmetries, 
namely the {\em parity}
\begin{equation}
\hat{\cal P} : c_j \rightarrow c_{L-j},\quad \hat{\cal P}U =
U\hat{\cal P},\quad \hat{\cal P}^2 = \hat{1},
\label{eq:parity}
\end{equation}
and, for even size $L$, the particle-hole transformation
\begin{equation}
\hat{\cal R} : c_j \rightarrow (-1)^j c^\dagger_j,\quad
\hat{\cal R}U = U\hat{\cal R},\quad \hat{\cal R}^2 = \hat{1}.
\label{eq:parthole}
\end{equation}
Note on notation: Symbols wearing a {\em hat} $\hat{}$ denote linear 
transformations over the operator space of quantum observables.

\section{First Method: Direct time evolution and correlation functions}

For a fixed size $L$ and fixed number of fermions $N$, a unitary quantum
many-body map U (\ref{eq:mbmap}) acts over a Fock space of dimension 
\begin{equation}
{\cal N} = {L \choose N} = \frac{L!}{N! (L-N)!}.
\end{equation}
Dynamics of a given initial many-body state $\ket{\psi(0)}$
is given by a simple iteration of the Floquet map
\begin{equation}
\ket{\psi(m)} = U\ket{\psi(m-1)} = U^m\ket{\psi(0)}.
\end{equation}
Many body states $\ket{\psi}$ can be expanded in a complete
basis of the Fock space (of Slater determinants),
for which we may choose either the {\em position states}, 
labelled by sets of $N$ ordered integers $\vec{j}=(j_1,\ldots,j_N)$
\begin{equation}
\ket{\vec{j}} = c_{j_1} \cdots c_{j_N}\ket{0},
\quad 0\le j_1\le\ldots\le j_N < L,
\label{eq:posstat}
\end{equation}
or the {\em momentum states}, labeled by sets of $N$ ordered
integers $\vec{k}$
\begin{equation}
\ket{\vec{k}} = \tilde{c}_{k_1} \cdots \tilde{c}_{k_N}\ket{0},
\quad 0\le k_1\le\ldots\le k_N < L.
\label{eq:momstat}
\end{equation}
An important observation, implicitly made already in previous section
(\ref{eq:mbmap},\ref{eq:WW},\ref{eq:Tmom}),
is that kinetic propagator $\exp(-iT)$ is {\em diagonal} in momentum representation
while potential propagator $\exp(-iW)$ is {\em diagonal} in position representation
\begin{eqnarray}
U^T_{\vec{k},\vec{k}'} &:=& 
\bra{\vec{k}}\exp(-iT)\ket{\vec{k}'} \nonumber \\
&=& \delta_{\vec{k},\vec{k}'}
\exp\left(it \sum_{n=1}^N (\cos(s k_n)-1)\right),\\
U^W_{\vec{j},\vec{j}'} &:=&
\bra{\vec{j}}\exp(-iW)\ket{\vec{j}'} \nonumber \\
&=& \delta_{\vec{j},\vec{j}'}
\exp\left(-iV\sum_{n,n'=1}^N \delta_{j_{n'},j_{n}+1}\right).
\end{eqnarray}
Therefore, one may formulate a very efficient algorithm to perform
explicit time evolution of many-body states, provided that it is possible to switch 
between the two representations (\ref{eq:posstat}) and (\ref{eq:momstat}) 
as efficiently as in the problem of one kicked quantum particle $N=1$, e.g.
kicked rotor \cite{KR} or kicked Harper model \cite{KH}, 
namely by means of Fast Fourier Transformation (FFT) algorithm. 
Indeed, we succeeded to develop a fast algorithm 
which performs such anti-symmetrized multi-dimensional discrete 
Fourier transformation
\begin{equation} 
F_{\vec{j},\vec{k}} = \braket{\vec{j}}{\vec{k}}
\end{equation}
in roughly ${\cal N} \log_2 {\cal N}$ Floating Point Operations (FPO).
It is based on a factorization of $L-$site discrete Fourier 
transformation into the product of $\sim L \log_2 L$ 
2-site transformations parametrized with $2\times 2$ sub-matrices
$(\alpha,\beta;\gamma,\delta)_{j j^\prime}$, which are successively 
applied to creation operators, 
$(c^\dagger_j,c^\dagger_{j^\prime})\leftarrow
(\alpha c^\dagger_j + \beta c^\dagger_{j^\prime},
 \gamma c^\dagger_j + \delta c^\dagger_{j^\prime})$,
in all Slater determinants $\Pi_n c^\dagger_{j_n}\ket{0}$ 
which contain a particle at sites $j$ or $j^\prime$.
(One should be careful in dealing with
fermionic signs of Slater determinants when sorting the
factors in the product $\Pi_n c^\dagger_{j_n}\ket{0}$). 
In case when $L=2^p$, factorization of FFT to a chain of 2-site 
(in such case unitary) transformations is easily deduced by 
inspecting a conventional FFT algorithm (such as the one implemented in
\cite{Recipes}),
while for more general lattice sizes (we have so far implemented our
Fermionic FFT (FFFT) algorithm also for $L=10,12,15,20,24,30,40$)
we factorized the optimal schemes developed by Winograd \cite{Winograd}.
Our FFFT algorithm requires almost no extra storage apart from a 
vector of ${\cal N}$ c-numbers where quantum many-body state is 
stored.
Therefore, the map (\ref{eq:mbmap}) is iterated on a vector
$\psi_{\vec{k}}(m) = \braket{\ve{k}}{\psi(m)}$, using the matrix 
composition
\begin{equation}
U = F^* U^W F U^T
\label{eq:decomp}
\end{equation}
in roughly $2{\cal N}\log_2 {\cal N}$ FPO per time step
which is by far superior to `brute-force' methods based on 
complete diagonalization of
$U$ and expansion of time-evolving state $\ket{\psi(m)}$ in 
terms of eigenstates of $U$.

Let us now consider time autocorrelation functions of
two `generic' observables, namely the current $J$ and 
rescaled traceless kinetic energy $T'$
\begin{eqnarray}
C^L_J(m) &:=& \frac{1}{L} \ave{J(0) J(m)}^L_\rho \label{eq:cor1} \\
C^L_T(m) &:=& \frac{1}{L} \ave{T'(0) T'(m)}^L_\rho,\quad
T' = \frac{1}{t} T - N \nonumber
\end{eqnarray}
$J(m)={U^\dagger}^m J U^m$,
$T'(m)={U^\dagger}^m T' U^m$.
Note that $\ave{J}^L_\rho = \ave{T'}^L_\rho = 0$.
These two observables belong to different symmetry classes
with respect to the parity operation
\begin{equation}
\hat{\cal P} J = - J\hat{\cal P},\quad
\hat{\cal P} T' = T'\hat{\cal P},
\end{equation}
so we choose both to check whether many-body dynamics can
substantially depend of the symmetry class w.r.t. `reflection' 
symmetry. Conveniently, both observables, $J$ and $T'$, are 
diagonal in momentum representation, 
\begin{eqnarray}
J\ket{\vec{k}} = J_{\vec{k}}\ket{\vec{k}},\quad
J_{\vec{k}} = \sum_{n=1}^N \sin(s k_n),\\
T'\ket{\vec{k}} = T'_{\vec{k}}\ket{\vec{k}},\quad
T^{\prime}_{\vec{k}} =-\sum_{n=1}^N \cos(s k_n),
\end{eqnarray}
so the time auto-correlation functions
can be computed from time evolution of a (complete) set of
${\cal N}'$ ($={\cal N}$) initial momentum states $\vec{k}'$
\begin{equation}
C_A(m) = \frac{1}{L{\cal N}'} 
\sum^{\prime}_{\vec{k}'} A_{\vec{k}'} \sum_{\vec{k}}
A_{\vec{k}} p_{\vec{k}\vec{k}'}(m)
\label{eq:corr}
\end{equation}
where $A$ is any observable which is diagonal in momentum basis
(here either $J$ or $T'$) and 
\begin{equation}
p_{\vec{k}\vec{k}'}(m) = |\braket{\vec{k}}{\psi(m)}|^2 =
|\bra{\vec{k}}U^m\ket{\vec{k}'}|^2.
\label{eq:p}
\end{equation}
When dimensionality ${\cal N}$ becomes prohibitively large we suggest
to estimate the microcanonical averages (\ref{eq:cor1})
by taking a smaller $1 \ll {\cal N'} \ll {\cal N}$ but uniformly
random sample of ${\cal N'}$ initial states $\ket{\vec{k}'}$,
$\ave{.} = 
\frac{1}{\cal N'}\sum^{\prime}_{\vec{k'}}\bra{\vec{k'}}.\ket{\vec{k'}}
(1 + {\cal O}(1/\sqrt{\cal N'})).$ 
Therefore, numerical computation of correlation function $C_A(m)$ for 
$m=1,\ldots,M$ can be performed in 
$\sim (2 M {\cal N'}{\cal N}\log_2{\cal N})/L$ FPO.
Reduction for a factor $1/L$ w.r.t. naive FPO count is
due to translational symmetry (\ref{eq:transs}), since
one can simultaneously simulate the dynamics of $L$ 
different states with different values of the conserved
total momentum $K=\sum_{n=1}^N k'_n \pmod{L}$.

Let us for the time being fix the density of particles
$\rho=N/L=1/4$. We have performed extensive numerical computations
of time-correlation functions (by means of explicit time evolution
(\ref{eq:corr})), for sizes $L=8,12,16,20,24,32$ 
(at $L=32$ the dimensionality of Fock subspace is 
${\cal N} = 10518300$), and systematically scanned the
parameter space $(t,V)$.
We have clearly identified two regimes where we were able
to probe the TL, i.e. where time-correlation functions
turned out to be stable against the variation of the system
size $L$: 
\\\\
(i) Quantum ergodic and mixing regime for sufficiently large
value of parameter $t$ and for any value of parameter $V$ (away from
`integrable axis' $V=0 \pmod{2\pi}$).
In this regime, time correlation functions are rapidly decreasing
(Case $t=V=4$ is shown in Fig.1).
However, for a finite size $L$ the quantum system is almost never mixing,
so correlation functions saturate, on a time scale $\mu(L)$, to a 
small but finite value of the stiffness
\begin{equation}
D^L_A = \lim_{M\rightarrow\infty}\frac{1}{2M+1}\sum_{m=-M}^M C^L_A(m)
\label{eq:stifA}
\end{equation}
(Here $A=J$ or $T'$.) In order to avoid transient behavior at small times $m$
and incorporating the time reversal symmetry, $C^L_A(m) = C^L_A(-m)$,
the time averages like (\ref{eq:stifA}) have been numerically estimated as
$D^L_A = \frac{1}{M'+1}\sum_{m=M'}^{2M'} C^L_A(m)$. 
The sufficiently large averaging time scale $\{M'\ldots 2M'\}=\{30 \ldots 60\}$ for 
all sizes $L\le 32$ (and for most values of parameters $t,V$) has been determined 
by direct inspection of the correlation functions (see Fig.1). 
In Fig.2 we plot correlation functions $C^L_J(m)$ for $t=V=4$ in
semi-log scale, and show that, as the size $L$ is increased, 
the saturation (Thouless) time scale $\mu(L)$ increases, roughly as 
$\mu(L) \sim L$. So, the Thouless time $\mu(L)$ clearly diverges in TL.
Furthermore, Fig.2 gives clear numerical evidence 
(further supported by results shown in  Figs.4,5) of the 
{\em exponential} decay of
time correlation functions in TL (or for times smaller than $\mu(L)$ in a 
system of finite size $L$)
\begin{equation}
C_A(m) \propto \exp(-\lambda_A m),\quad m \gg 1.
\label{eq:expmix}
\end{equation}
Henceforth, the stiffness should also vanish exponentially 
$D^L_A \sim \exp(-\lambda_A L)$ as one approaches TL
(Such behavior has been observed also in Ref.\cite{Prelovsek} indicating
exponential mixing in the system sudied there \cite{Prelovsek}).
Indeed, in Fig.3 we examine $1/L$ scaling of the charge stiffness 
$D^L_J$ which is (here shown for $t=V=4$ and $t=V=2$) a clear 
indication of ergodic and mixing behavior in TL, $D^\infty_J=0$. 
In this regime, in TL, Kubo conductivity
$\sigma'(\omega=0) = \half\beta\sum_{m=-\infty}^\infty
C_J(m)$ is finite, $\sigma'(0)<\infty$, and the transport is 
dissipative. Further, as shown in \cite{Prosen1},
time averaged current of arbitrary initial momentum state
$\ket{\vec{k}'}$ averages to zero $\bar{J}_{\vec{k}'}=
\lim_{M\rightarrow\infty}(1/M)\sum_{m=1}^M
\bra{\vec{k}'} J(m)\ket{\vec{k}'} = 0$,
and aribtrary initial momentum state $\ket{\vec{k}'}$
explores entire accessible Fock space, i.e. 
$\bra{\vec{k}}U^m\ket{\vec{k}'}$ are uniformly Gaussian pseudo-random
numbers when the discrete time $m$ is sufficiently large, say larger than 
the quantum mixing time.
\\\\
\begin{figure}[htbp]
\hbox{\hspace{-0.1in}\vbox{
\hbox{
\leavevmode
\epsfxsize=3.6in
\epsfbox{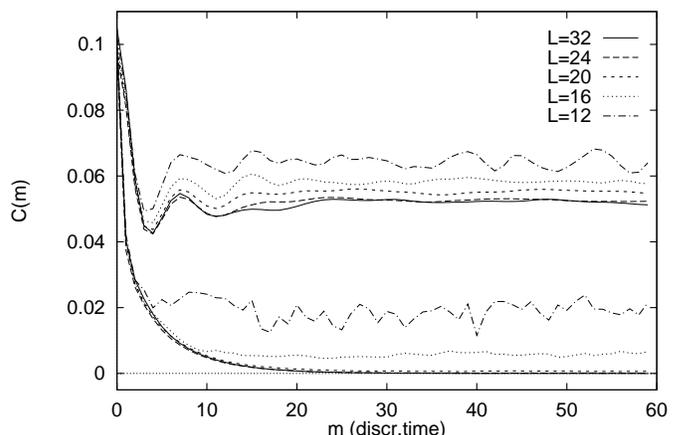}}
}}
\caption{Current autocorrelation function $C_J(m)$ against discrete time
$m$ for quantum ergodic ($t=V=4$, lower set of curves for various sizes 
$L$) and intermediate regime ($t=V=1$, upper set of curves) with
density $\rho=\frac{1}{4}$. Averaging over entire Fock space is performed, 
${\cal N}^\prime={\cal N}$, for $L\le 20$, whereas random samples of 
${\cal N}^\prime=12000$, and ${\cal N}^\prime=800$ initial states 
have been used for $L=24$, and $L=32$, respectively.}
\label{fig:s3f1}
\end{figure}

\noindent (ii) Non-ergodic and Non-mixing regime
for parameter $t\sim 1$ (or smaller) and any value of parameter 
$V$. Here time correlation functions $C_A(m)$ do not
decay to zero but saturate, around a constant typically non-vanishing 
and positive value of the stiffness $D^L_A > 0$ (\ref{eq:stifA}), 
on a short time scale which {\em does not} depend on the size $L$
(for sufficently large sizes $L$). In Fig.1 we plot time correlation 
functions $C_J(m)$
for $t=V=1$. Please observe indeed very weak dependence on the size $L$.
In Fig.3 we also show $1/L$ scaling of charge stiffness $D^L_J$
for the cases $t=V=1$ and $t=1,V=2$ which clearly indicate
a finite extrapolated (to $1/L=0$) thermodynamic value of
the stiffness.
This should be considered as evidence of non-mixing and non-ergodic 
behavior in TL. Since in this parameter ranges KtV model is
also non-integrable, we sometimes refer to this regime as an {\em
intermediate quantum dynamics}. This behavior corresponds to 
ideal, ballistic transport with infinite Kubo conductivity 
$\sigma=\infty$.
Furthermore, in \cite{Prosen1} it has been shown that in this
intermediate regime the time averaged (persistent) current is 
{\em non-vanishing} and {\em proportional} to the initial current 
$J_{\vec{k}'}$, $\bar{J}_{\vec{k}'} = \alpha J_{\vec{k}'},\;
\alpha=2 D_J/[\rho (1-\rho)]$ (which is the most direct probe of
ideal, ballistic transport),
and that an arbitrary time-evolving initial momentum state 
$U^m\ket{\vec{k}'}$ remain strongly localized in a
non-trivial subregion of dynamically accessible Fock space.
\\\\
\begin{figure}[htbp]
\hbox{\hspace{-0.1in}\vbox{
\hbox{
\leavevmode
\epsfxsize=3.6in
\epsfbox{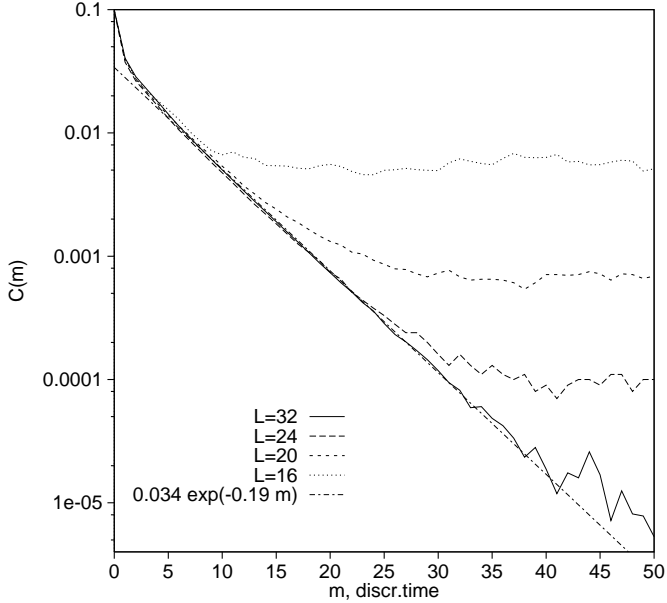}}
}}
\caption{
The lower set of curves ($t=V=4$) of Fig.1 in semi-log scale.
To emphasize the exponential correlation decay we plot also
the best exponential fit to the tail of $C^{L=32}_J(m)$
(dash-dotted line). 
}
\label{fig:s3f2}
\end{figure}
\begin{figure}[htbp]
\hbox{\hspace{-0.1in}\vbox{
\hbox{
\leavevmode
\epsfxsize=3.6in
\epsfbox{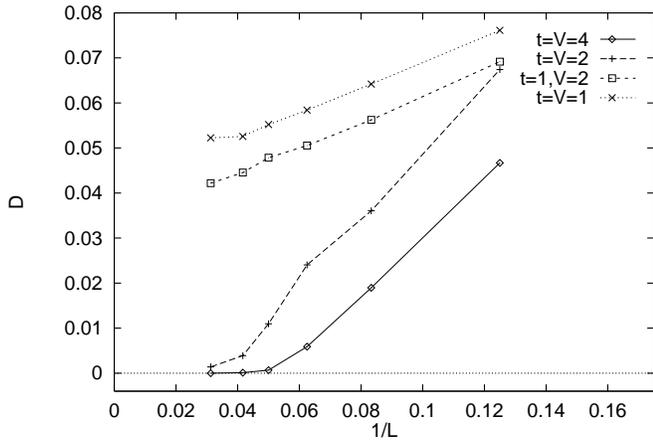}}
}}
\caption{
Stiffness $D_J$ vs. $1/L$
at constant density $\rho=\frac{1}{4}$ and for different 
values of control parameters in quantum mixing/ergodic, $t=V=4$ 
and $t=V=2$, and intermediate, $t=1,V=2$ and $t=V=1$, regime.
Other parameters are the same as in Fig.1
}
\label{fig:s3f3}
\end{figure}
\noindent
One may use a charge stiffness of an infinite system $D^\infty_J$
as an order parameter controlling the {\em dynamical
phase transition} from a disordered phase (quantum ergodic/mixing
dynamics) characterized by $D^\infty_J = 0$ to an ordered phase 
(non-ergodic/non-mixing dynamics) characterized by $D^\infty_J > 0$.
The transition point is characterized by diverging correlation
time (or mixing time) scale, $\lambda_J^{-1}$, which diverges when
one approaches the transition from above, say, with parameter 
$t$ decreasing towards certain critical curve $t_c(V)$.
Of course, in the ordered phase, $t < t_c(V)$, $D^\infty_J>0$, 
the time-correlations have infinite range, $C_A(\infty)\neq 0$.
The transition is illustrated in Figs.4,5 by plotting correlation functions
for both observables, $C_J(m)$ (Fig.4) and $C_T(m)$ (Fig.5), for 
different values of parameter $t$ and fixed parameter $V=2$.
En estimate of the critical parameter here is $1.4 < t_c(V=2) < 1.5$. 
Observe the exponential decay of correlations in all cases, 
except possibly at very small times where other, smaller
time scales may become important.

\begin{figure}[htbp]
\hbox{\hspace{-0.1in}\vbox{
\hbox{
\leavevmode
\epsfxsize=3.6in
\epsfbox{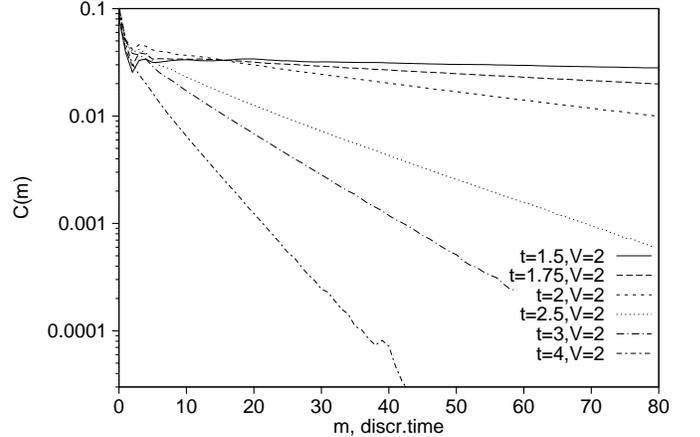}}
}}
\caption{
Current auto-correlation functions $C^{L=32}_J(m)$ for
different values of parameter $t$ (see legend) and fixed 
value of parameter $V$. $\rho=1/4$.
}
\label{fig:s3f4}
\end{figure}

\begin{figure}[htbp]
\hbox{\hspace{-0.1in}\vbox{
\hbox{
\leavevmode
\epsfxsize=3.6in
\epsfbox{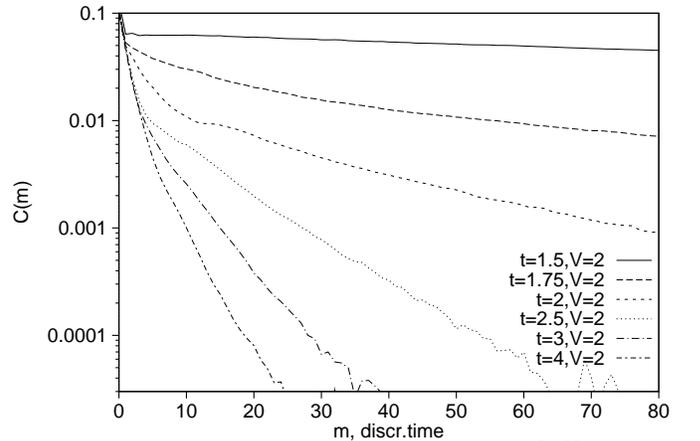}}
}}
\caption{
Kinetic auto-correlation functions $C^{L=32}_T(m)$ for
different values of parameter $t$ (see legend) and fixed 
value of parameter $V$. $\rho=1/4$. Note that the same
scale is used than in previous Fig.4.
}
\label{fig:s3f5}
\end{figure}

\section{Second Method: Exact diagonalization of stationary
problem of finite size}

In a more complete but brute-force approach one may try to exactly 
diagonalize the matrix of a one-period evolution (Floquet) propagator $U$ for 
a finite size $L$ and compute interesting
dynamical quantities, such as conductivity $\sigma'(\omega)$
and stiffness $D^L_A$ directly from the spectrum $\{\varphi_n\}$
and the set of eigenstates $\{\ket{n}\}$ of KtV map $U$.
\begin{equation}
U\ket{n} = \exp(-i\varphi_n)\ket{n},\quad n=1\ldots{\cal N}.
\end{equation}
Again, it is easiest to work in the momentum basis (\ref{eq:momstat}) and to use 
the translational symmetry to decompose the matrix $U_{\vec{k},\vec{k}'}$ into
blocks (with fixed value of the total quasi-momentum $K$) of 
dimension ${\cal N}_K \approx {\cal N}/L$.
Only for blocks with $K=0$ and $K=L/2$ (if $L$ is even) 
the parity operation $\hat{\cal P}$ (\ref{eq:parity}) commutes with the translation 
$S$ (\ref{eq:transs}), and may be used to
further reduce the dimensionality of irreducible block by factor 2.
The matrix $U_{\vec{k},\vec{k}'}$ (for fixed $K$) 
has been computed, by means of a decomposition (\ref{eq:decomp}) and FFFT 
algorithm, in roughly $({\cal N}/L)^2\log_2{\cal N}$ FPO, and
further diagonalized by means of standard routines
in roughly $({\cal N}/L)^3$ FPO giving a set of quasi-energies $\{\varphi_n\}$ 
and eigenstates $\braket{\vec{k}}{n}$.

\subsection{Spectral statistics}

In the so called Quantum Chaology of simple (few) body non-integrable 
system there is a famous Conjectue of Bohigas, Giannoni and Schmit \cite{BGS}, 
supported by numerous numerical \cite{Bohigas} and theoretical arguments 
\cite{Alt}, claiming that hard chaos (ergodicity, mixing and positive
Lyapunov exponents) of a classical counterpart results in a universal 
statistical properties of system's (quasi)energy spectrum given by the 
appropriate ensemble of random matrices\cite{Mehta}. On the other hand, 
integrable classical 
classical dynamics results in universal Poissonian statistics
of (locally) uncorrelated (quasi)-energy levels \cite{BerryTabor}.
Intermediate statistics, which are neither RMT nor Poissonian, 
are found \cite{BR84,PR94} for systems whose classical
dynamics is intermediate (mixed) with regular and chaotic motion coexisting
in phase space. The connection between integrability/non-integrability and
statistics has recently been investigated also in few well known examples of 
non-linear many-body systems (correlated fermions or interacting spin chains) 
which do not possess a well defined classical limit \cite{fRMT1}. 
It has been shown that quantum integrability, or (strong) non-integrability, 
of the quantum many-body model again correspond to Poissonian, or RMT, 
behavior of level statistics, respectively.
No attempt has been made there \cite{fRMT1}, however, to understand
the intermediate situation, or the thermodynamic limit.

Inspired by Quantum Chaos we have analyzed the statistical properties of 
the quasi-energy spectrum $\{\varphi_n\}$ of KtV model and searched for 
signatures of ergodicity and mixing of the underlying quantum many-body 
dynamics in TL. For comparison with other results, the density will be again
fixed to $\rho=1/4$ in the numerical presentation which follows.

First, we have analyzed the common short-range statistic, namely
the integrated (cumulative) nearest neighbor level spacing distribution,
$W(S)$, giving the probability that a random normalized spacing between
two adjacent eigenphases 
$S_n=\frac{{\cal N}}{2\pi L}(\varphi_{n+1}-\varphi_n)$
is smaller than $S$
\begin{equation}
W(S) = \frac{1}{\cal N'}\sum^\prime_{n} \theta(S - S_n).
\label{eq:W}
\end{equation}
\begin{figure}[htbp]
\hbox{\hspace{-0.1in}\vbox{
\hbox{
\leavevmode
\epsfxsize=3.6in
\epsfbox{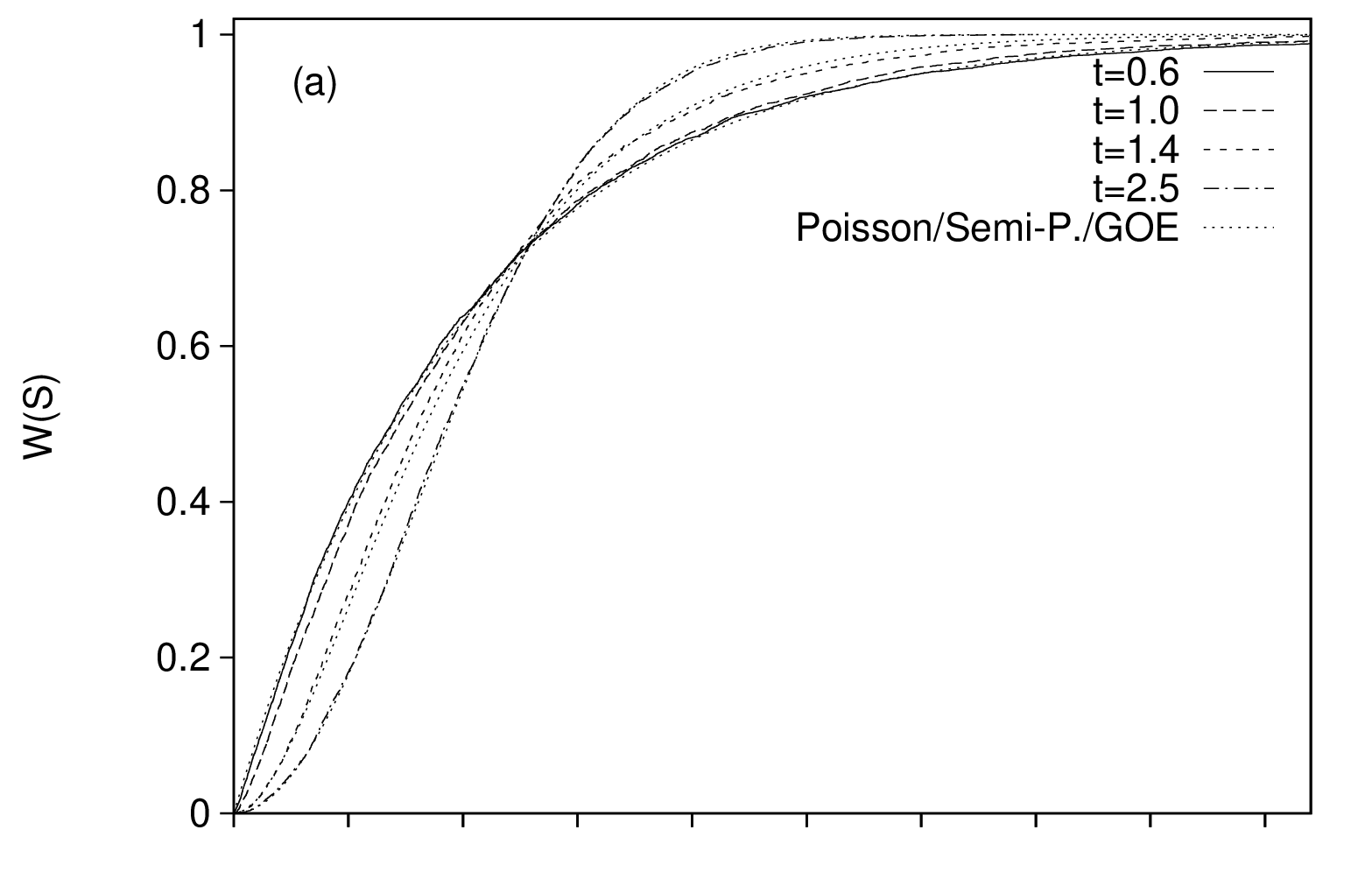}}
\vspace{-1.62in}
\hbox{\hspace{0.93in}
\vbox{
\leavevmode
\epsfxsize=2.45in
\epsfbox{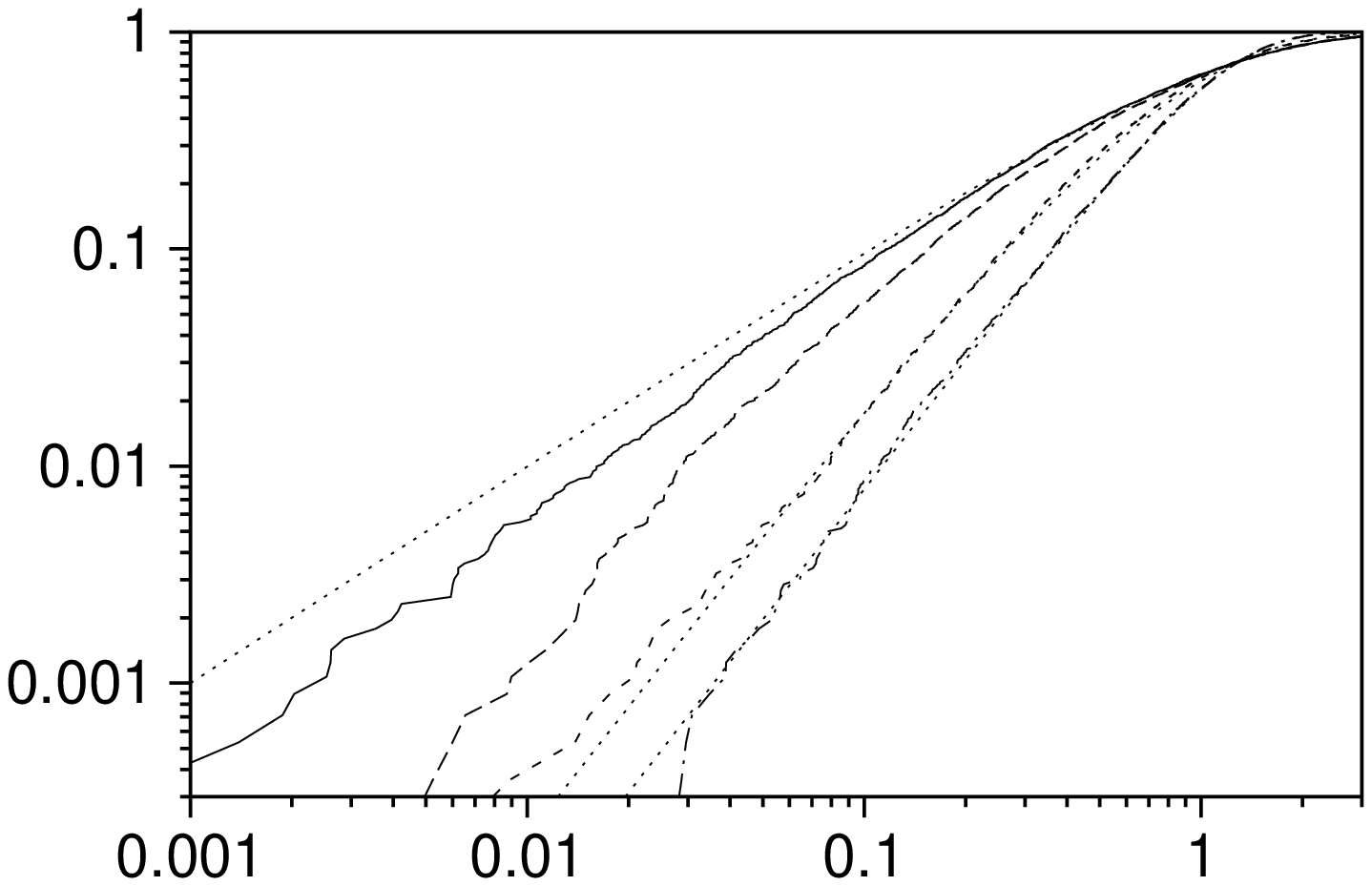}}
\vspace{0.4in}
}}}
\vspace{-0.33in}
\hbox{\hspace{-0.1in}\vbox{
\hbox{
\leavevmode
\epsfxsize=3.6in
\epsfbox{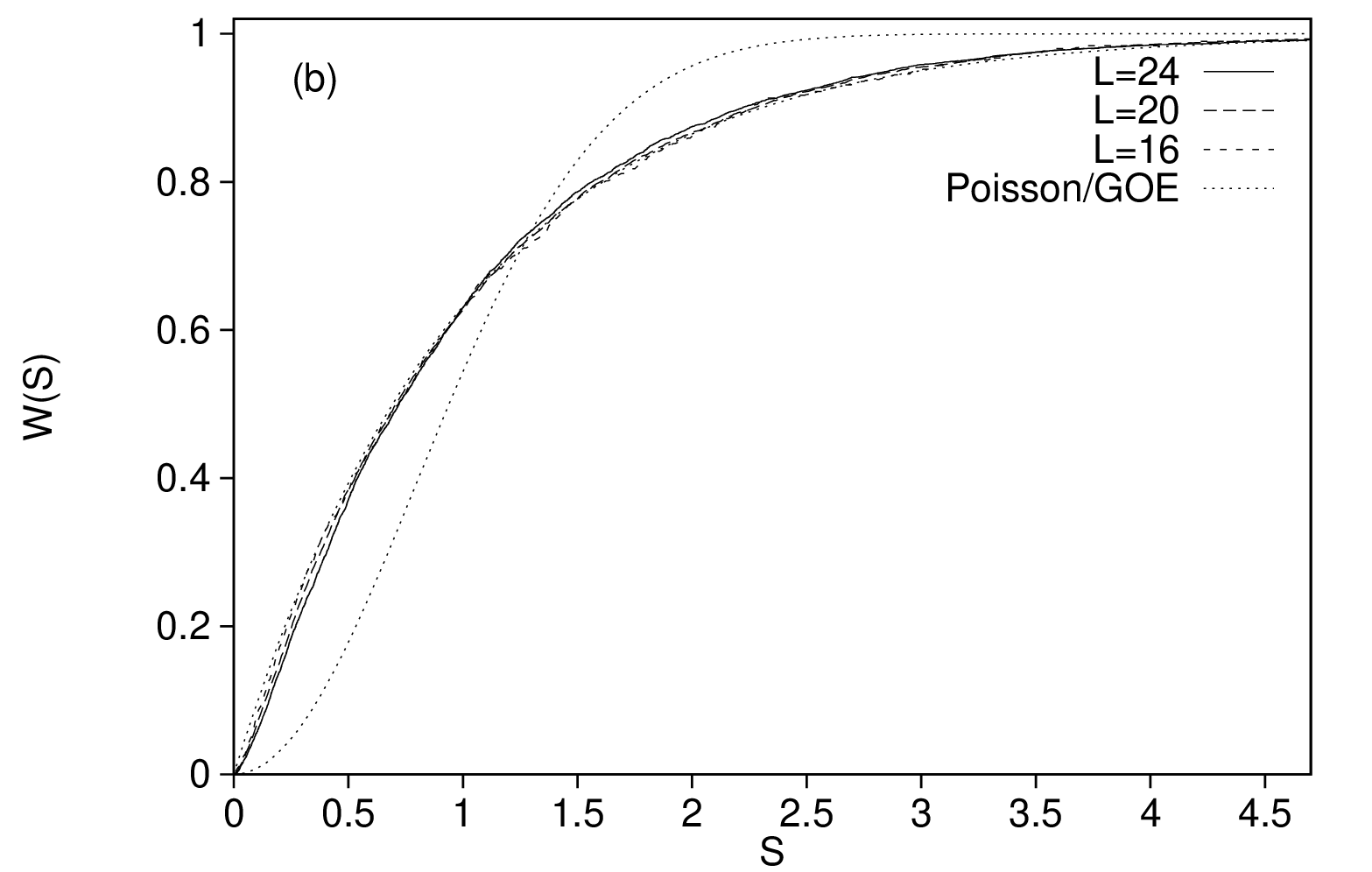}}
\vspace{-1.69in}
\hbox{\hspace{0.93in}
\vbox{
\leavevmode
\epsfxsize=2.45in
\epsfbox{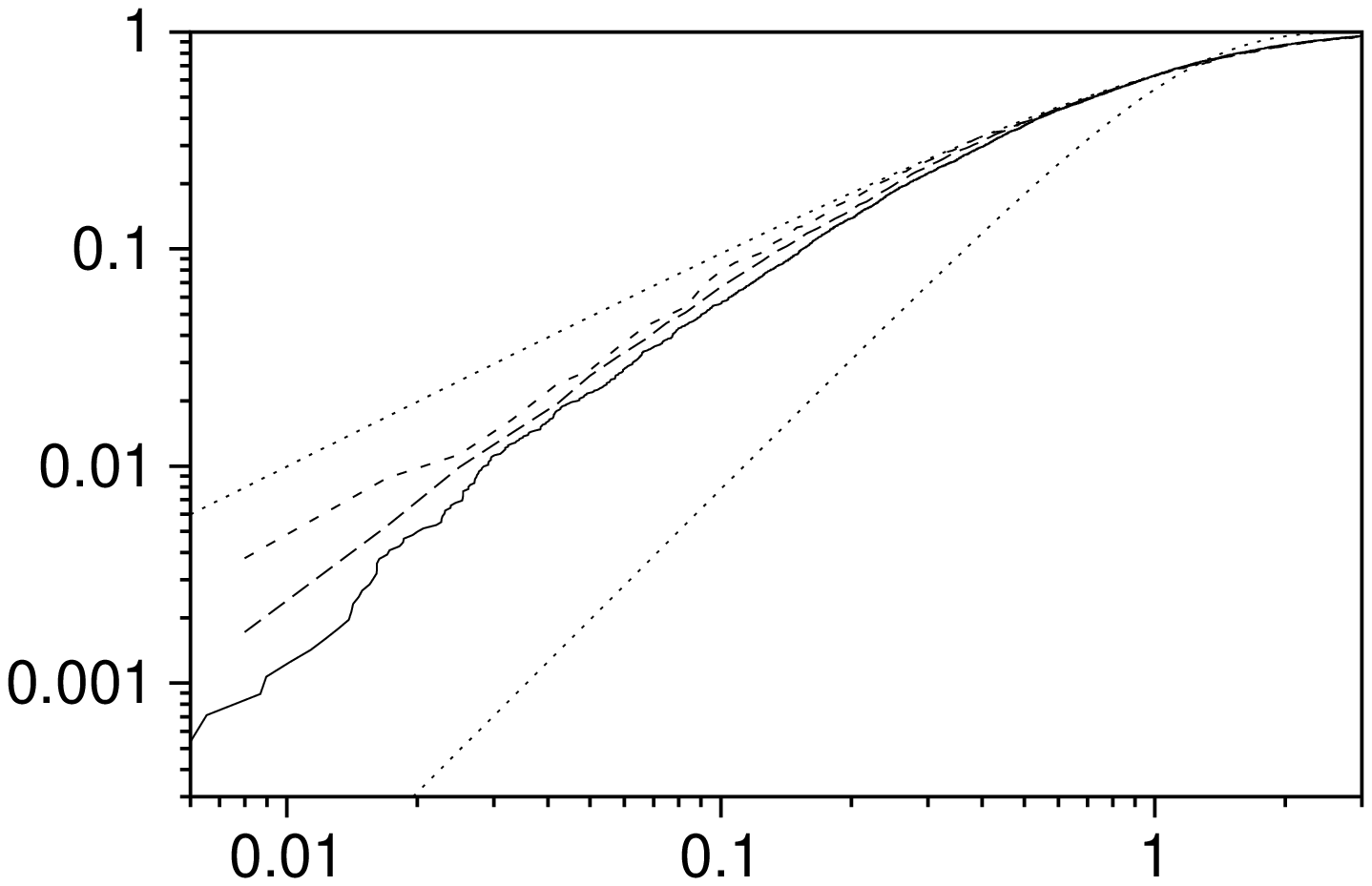}}
\vspace{0.4in}
}}}
\caption{
Cumulative quasi-level spacing distributions $W(S)$ for quarter-filled
$\rho=1/4$ KtV model. In (a) we plot $W(S)$ for several different
values of parameter $t$ and fixed parameter $V$ (see legend) covering the
mixing/non-mixing transition, and for maximal computable size $L=24$.
With dotted curves we plot for comparison the theoretical, Poissonian, 
Semi-Poissonian, and Wigner (COE), distributions.
In (b) we show $W(S)$ for fixed
kick parameters $t=1,V=2$, and for different sizes $L=16,20,24$.
In the insets we plot the same objects in log-log scale to emphasize
the small spacing behavior.
Please observe the trend towards linear repulsion (quadratic for 
$W(S\rightarrow 0) \propto S^2$), even in intermediate regime.
}
\label{fig:6}
\end{figure}

\noindent
For moderate values of the size $L\le 20$, the average in (\ref{eq:W}) 
has been computed over all ${\cal N}'={\cal N}$ states from all $L$ blocks 
(symmetry classes) of constant quasi-momentum $K$.
(Note that blocks for $K=K'$ and $K=L-K'$ are related by a parity
transformation (\ref{eq:parity}) and give identical (sub)spectra, so one
has to diagonalize only $[L/2+1]$ different blocks of matrix 
$U_{\vec{k},\vec{k}'}$.) 
However, for larger size $L=24$
we have already ${\cal N}_K \approx 5608$, 
so we averaged only over one class of fixed quasi-momentum, namely $K=1$,
${\cal N}' = {\cal N}_1 \approx {\cal N}/L$.
(We have carefully checked that the statistical properties of
partial subspectra are independent of the symmetry class labeled by 
quasi-momentum $K$.)

In Fig.6a we show $W(S)$ for size $L=24$ and several different 
values of parameter $t$ (and fixed $V=2$) covering the transition from 
non-ergodic to ergodic/mixing quantum dynamics.
We find almost Poissonian behavior $W_{\rm P}(S)= 1-\exp(-S)$ for small $t$ 
and excellent RMT behavior $W_{\rm COE}(S) = 1-\exp(-\pi S^2/4)$ 
(Wigner surmise approximating the statistics of the infinitely-dimensional 
Circular Orthogonal Ensemble COE
\cite{Mehta}, due to time-reversal symmetry) for $t > t_c(V)$. 
In the (more interesting) region of intermediate dynamics 
$1 \sim t < t_c(V)$ we find intermediate 
statistics interpolating between Poissonian and COE (see Fig.6).
Interestingly, level statistics close to the critical point (for $t=1.4,V=2$)
seems to be well captured by the so-called Semi-Poisson model
$W_{\rm SP}(S) = 1 - (1+2S)\exp(-2S)$ \cite{Bogomolny} which has been recently 
used to model the critical level statistics of 3D Anderson model
\cite{Montambaux}. Since it is impossible to make statements about TL of 
level statistics based on results for a fixed size $L=24$, we show in 
Fig.6b the dependence of $W(S)$ on the size $L$
for fixed parameters $t=1,V=2$ (in the regime of non-ergodic dynamics). 
Although the intermediate $W(S)$ statistic is closer to 
Poissonian than to COE, it moves towards COE as we approach TL (increase $L$).
This fact eliminates possible fears of accidental
integrability of KtV model in the claimed intermediate regime.

Second, we have analyzed the long-range spectral statistics, namely the
number variance
\begin{equation}
\Sigma^2(S) = \ave{n(S)^2} - S^2,
\end{equation}
giving the variance of the number $n(S)$ of normalized (unfolded) levels
$\frac{\cal N}{2\pi L} \varphi_n$ in a randomly chosen interval of length $S$
(Note that $\ave{n(S)}=S$.) For Poissonian and COE model we expect
$\Sigma^2_{\rm P}(S) = S$ and $\Sigma^2_{\rm COE}(S) \approx (2/\pi^2)\log(2\pi S)$,
respectively.
Here, one should note huge degeneracies in the integrable limit $t=0$ of Ising
model (which are quite-common in integrable quantum many-body models in general).
For small kick parameters $t$, we hence find stronger-than-Poissonian level 
clustering causing faster-than-linear growth of $\Sigma^2(S)$ (Fig.7a).
For finite size $L=24$, strong level clustering affects also long-range
statistics in the regime with mixing dynamics in TL, namely for
$t=2.5$ we find good agreement with COE statistics only for relatively small 
spectral ranges $S \le S_{\rm max} \sim 10^1$. It has been checked, however, 
that the agreement with COE improves to hold on longer quasi-energy ranges
($S_{\rm max}$ increases), as either the kick-parameter $t$ or the size $L$ are 
increased.
In the intermediate regime $1\sim t < t_c(V)$, number variance  
approaches that of an uncorrelated sequence,
$\Sigma^2(S) \sim S$, as we approach TL (see Fig.7b for case $t=1,V=2$).
However, for finite $L$, the phenomenon of saturation sets in \cite{Berry}, 
namely when the scaled energy range $S=S^*$ is of the order of the density 
of states
\begin{equation}
S^* = 0.5 {\cal N}/L,
\label{eq:satur}
\end{equation}
i.e. when the energy range $S^*$ becomes comparable to the length
of quasi-energy spectrum. Numerical factor $0.5$ in (\ref{eq:satur}) is of 
phenomenological origin.
Indeed for data of Fig.7, for $L=12,16,20,24$, the maxima of number variance
lie at $9,55,390,2800$, whereas theoretical values of $S^*$ (\ref{eq:satur})
are $9.1,56.9,387.5,2804$, respectively.

\begin{figure}[htbp]
\hbox{\hspace{-0.1in}\vbox{
\hbox{
\leavevmode
\epsfxsize=3.6in
\epsfbox{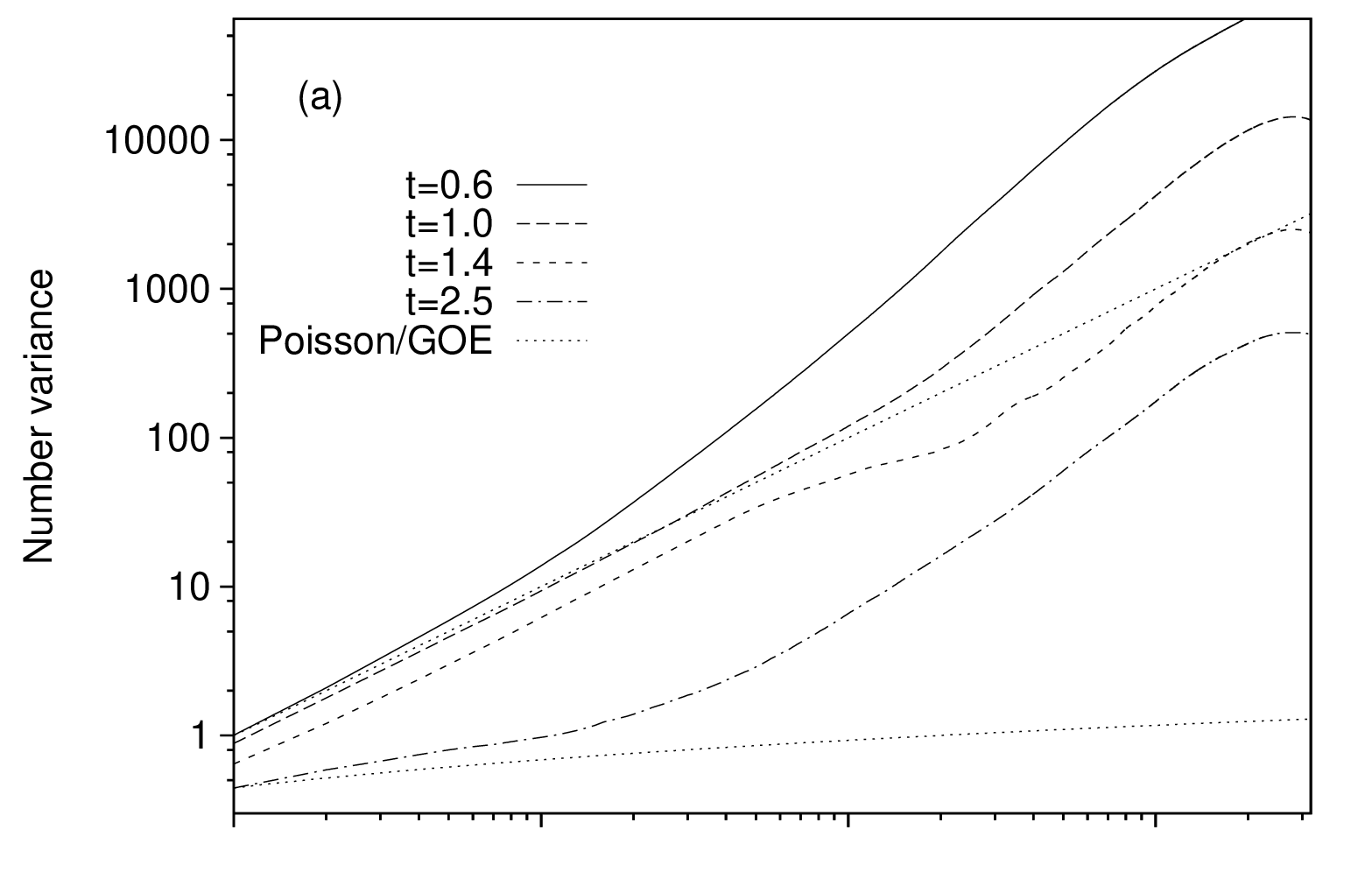}}
\vspace{-0.15in}
\hbox{
\leavevmode
\epsfxsize=3.6in
\epsfbox{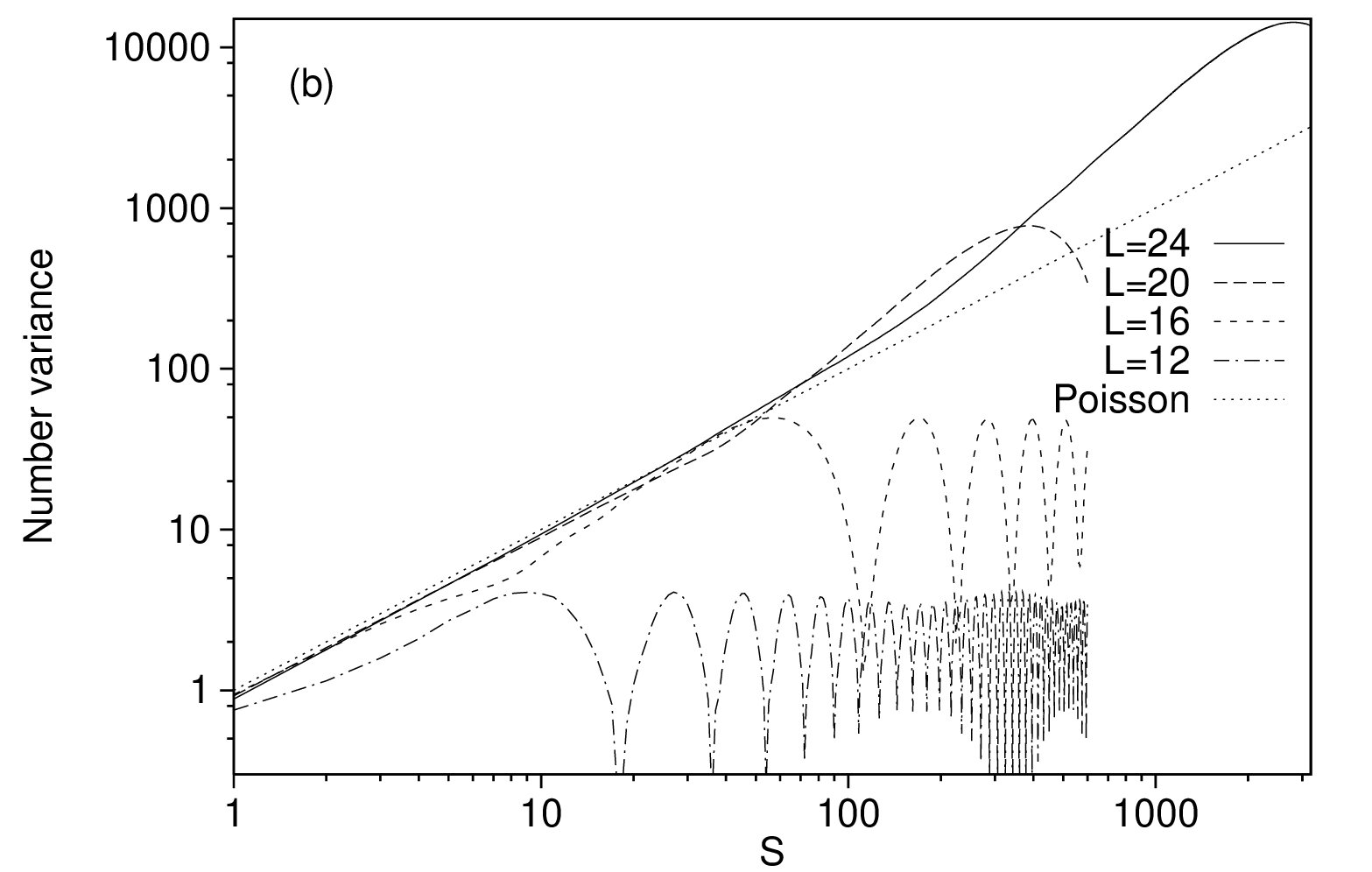}}
}}
\caption{Number variance $\Sigma^2(S)$ in log-log scale for exactly the same 
parameters (with an extra data in (b) for $L=12$) as in previous Figure 6.}
\label{fig:7}
\end{figure}

\subsection{Matrix elements and integrated conductance}

Knowing a complete set of eigenstates in momentum representation,
$\braket{\vec{k}}{n}$, it is easy to compute the matrix elements
of the current observable
\begin{equation}
\bra{n}J\ket{m} = \sum_{\vec{k}} J_{\vec{k}} 
\braket{n}{\vec{k}}\braket{\vec{k}}{m}.
\end{equation}
Again, one should make use of translational symmetry 
(\ref{eq:transs}), since $[S,J]=0$, to point out that matrix elements 
are non-vanishing only within a fixed quasi-momentum
block, $K_n \neq K_m \Rightarrow \bra{n}J\ket{m} = 0$. In order to obtain the
numerical results presented in this subsection we have averaged over the entire 
Fock space (all $K$), except again for $L=24,\rho=1/4$ where we have
averaged only over a block with quasi-momentum $K=1$. 
Real part of high-temperature electric conductivity (\ref{eq:kubo})
can be (for fixed size $L$) rewritten as
\begin{equation}
\sigma'(\omega) = \frac{\pi \beta }{L{\cal N}}
\sum_{n,m} |\bra{n}J\ket{m}|^2 
\delta_p({\textstyle{\frac{1}{2\pi}}}(\omega - \phi_m + \phi_n))
\end{equation}
In order to avoid ackward smoothing procedure and to
simplify the notation we introduce scaled integrated 
conductivity $I^L(\omega)$
\begin{eqnarray}
I^L(\omega) &=& \frac{2}{\beta}\int_{0}^{\omega+0} 
\sigma'(\nu) d\nu \nonumber\\
&=& \frac{1}{L{\cal N}}\sum_{n,m}
|\bra{n}J\ket{m}|^2 \theta(\omega + 0 - |\phi_m - \phi_n|'),
\end{eqnarray}
where $|\eta|' := \min\{|\eta|,2\pi - |\eta|\}$ and $\theta(x)$ is a Heaviside
step function.
Integrated conductivity $I^L(\omega)$ is a monotonically increasing function
on the frequency interval $\omega\in [0,\pi]$, starting from the 
charge stiffness 
\begin{equation}
I^L(0) = D^L_J 
\label{eq:Istif}
\end{equation}
and satisfying the sum-rule on the other end
\begin{equation}
I^L(\pi) = \oneL\ave{J^2}^L.
\end{equation}
Note that the current variance can be computed\cite{Prosen1}
\begin{equation}
\oneL\ave{J^2}^L = \frac{N(L-N)}{2L(L-1)} =
\half\rho(1-\rho) + {\cal O}\left(\oneL\right).
\end{equation}
In Fig.8 we plot the integrated conductivity $I^L(\omega)$
for different values of the kick parameter $t$ (fixed $V=2$) for constant
size $L=24$ and density $\rho=1/4$, showing the transition from ergodic
$D^L_J\approx 0$ to non-ergodic $D^L_J > 0$ dynamics, consistent 
with results of direct time evolution of Sec.III.
In Fig.9 we analyze the dependence of $I^L(\omega)$ on size $L$ for
fixed values of parameters in non-ergodic regime $t=1,V=2,\rho=1/4$. 
Note that for small frequencies $\omega$,
$I^L(\omega)$ is rougly constant over the frequency interval $0 \le \omega \le
\omega^L_*$, whose width is determined by the Thouless time of a finite system
$\mu(L)\sim L$, namely $\omega^L_*=2\pi/L$.
Note that the expression for stiffness (\ref{eq:Istif}) is not 
completely consistent with the correct definition (\ref{eq:stif}), 
since the time-limit $\tau\rightarrow\infty$ is implicit in (\ref{eq:Istif}) 
before TL of increasing $L$ can be considered, whereas the correct order
of limits is just the opposite. (This proves another advantage of the study of 
direct time-evolution of Sec.III over the more common frequency-domain 
approach presented here.)
\begin{figure}[htbp]
\hbox{\hspace{-0.1in}\vbox{
\hbox{
\leavevmode
\epsfxsize=3.6in
\epsfbox{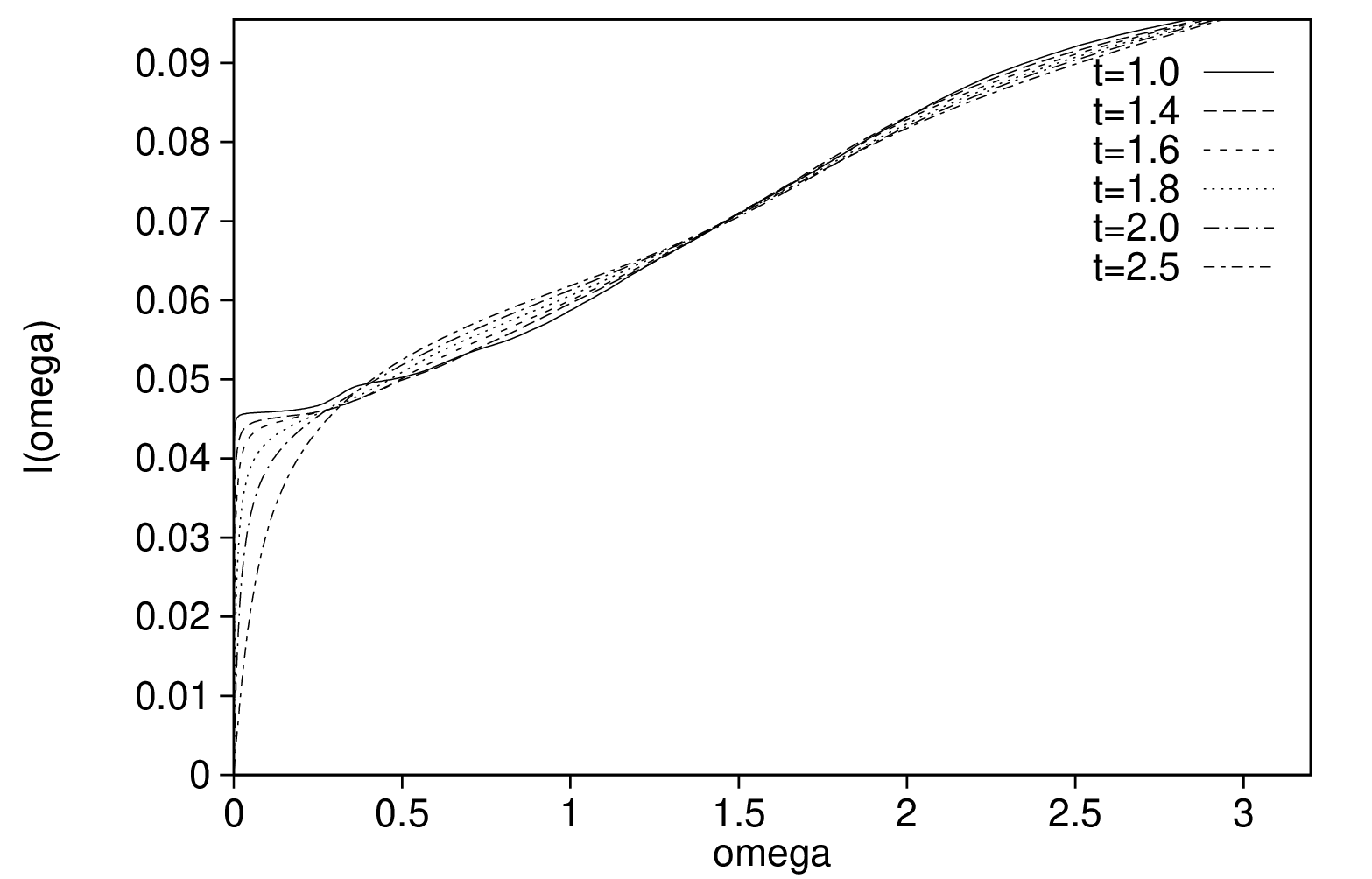}}
\vspace{-1.45in}
\hbox{\hspace{1.15in}
\vbox{
\leavevmode
\epsfxsize=2.2in
\epsfbox{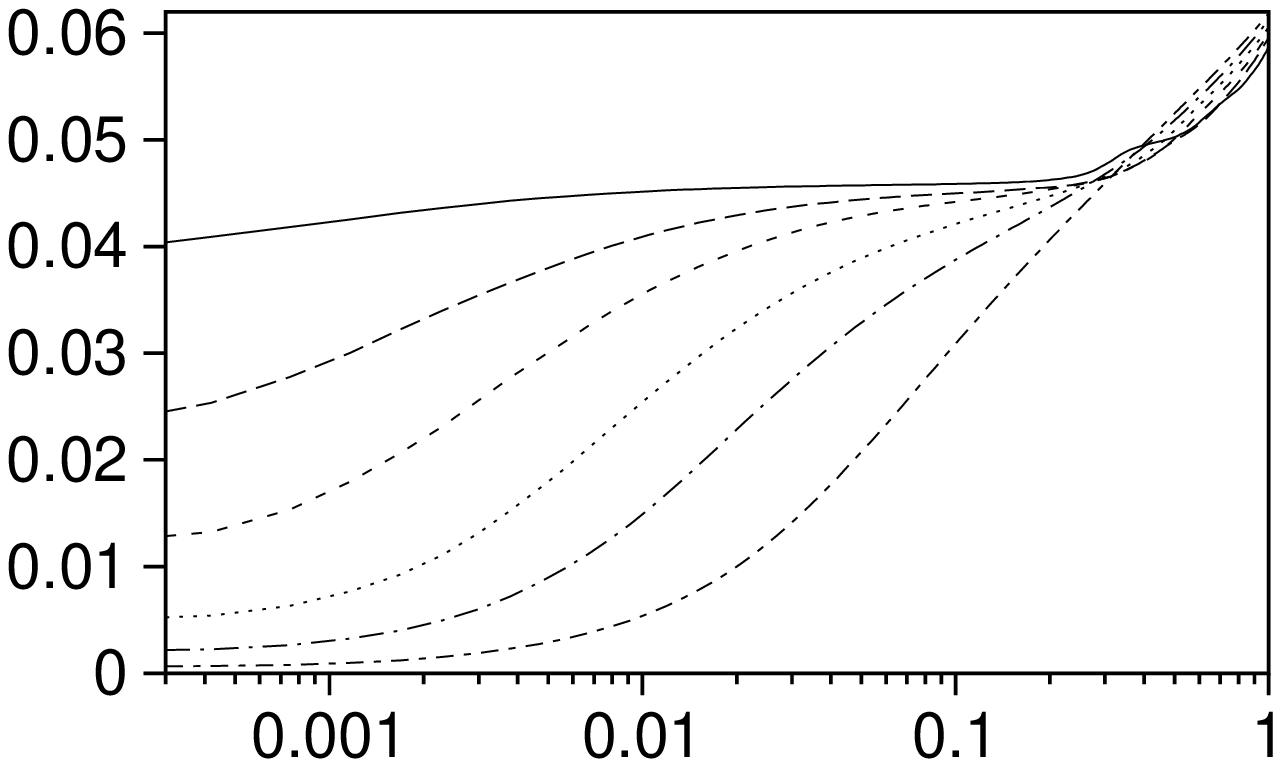}}
\vspace{0.4in}
}}}
\caption{We show the integrated conductance $I^L(\omega)$ for a cyclic 
chain of size $L=24$ and parameter $V=2$ and for several different values
of parameter $t$ (see legend).
In the inset we show the same plot in semi-log scale in order to
illustrate the zero freqency jump --- the charge stiffness $D^L_J$.}
\label{fig:8}
\end{figure}
\begin{figure}[htbp]
\hbox{\hspace{-0.1in}\vbox{
\hbox{
\leavevmode
\epsfxsize=3.6in
\epsfbox{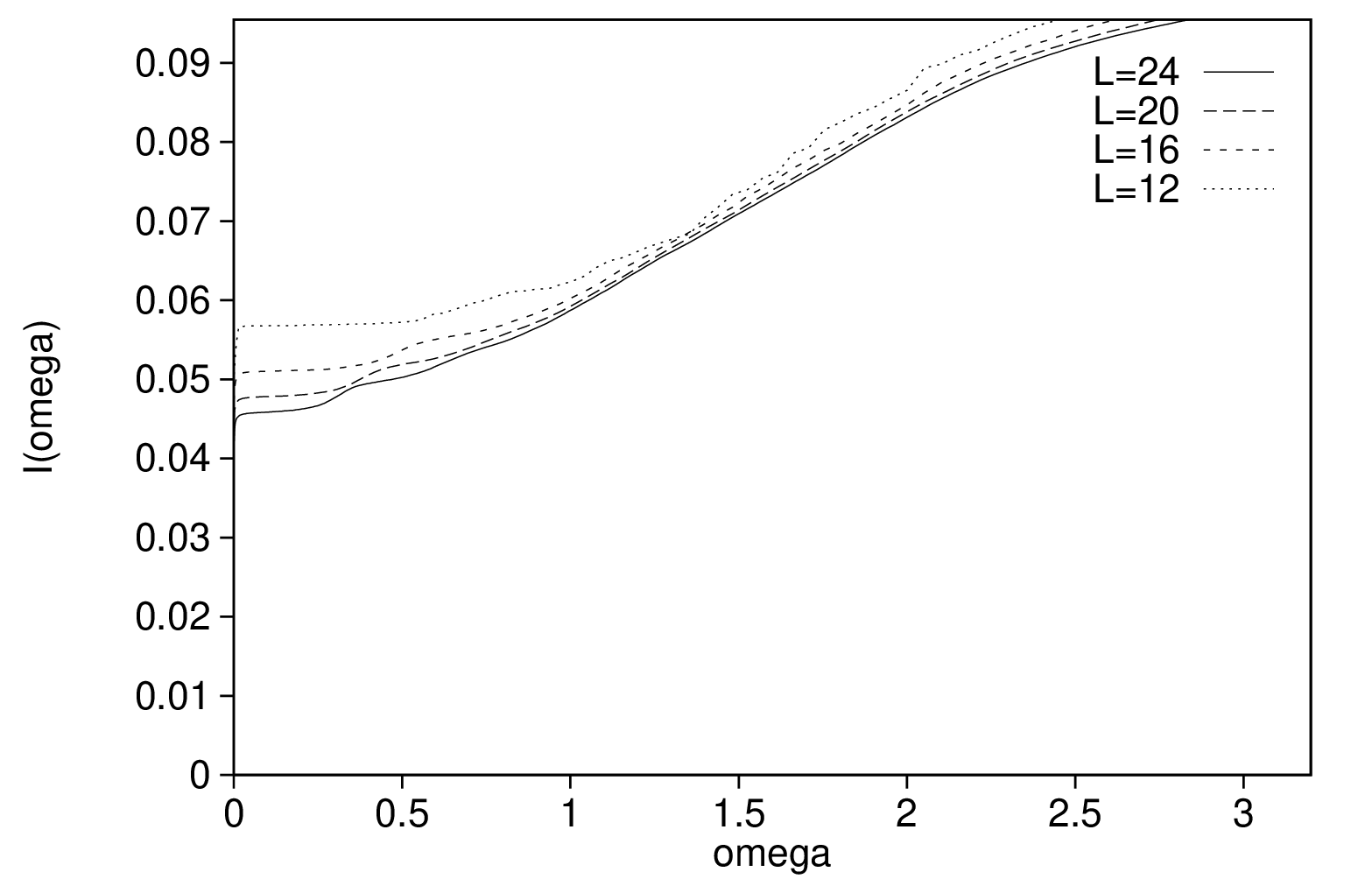}}
\vspace{-1.45in}
\hbox{\hspace{1.15in}
\vbox{
\leavevmode
\epsfxsize=2.2in
\epsfbox{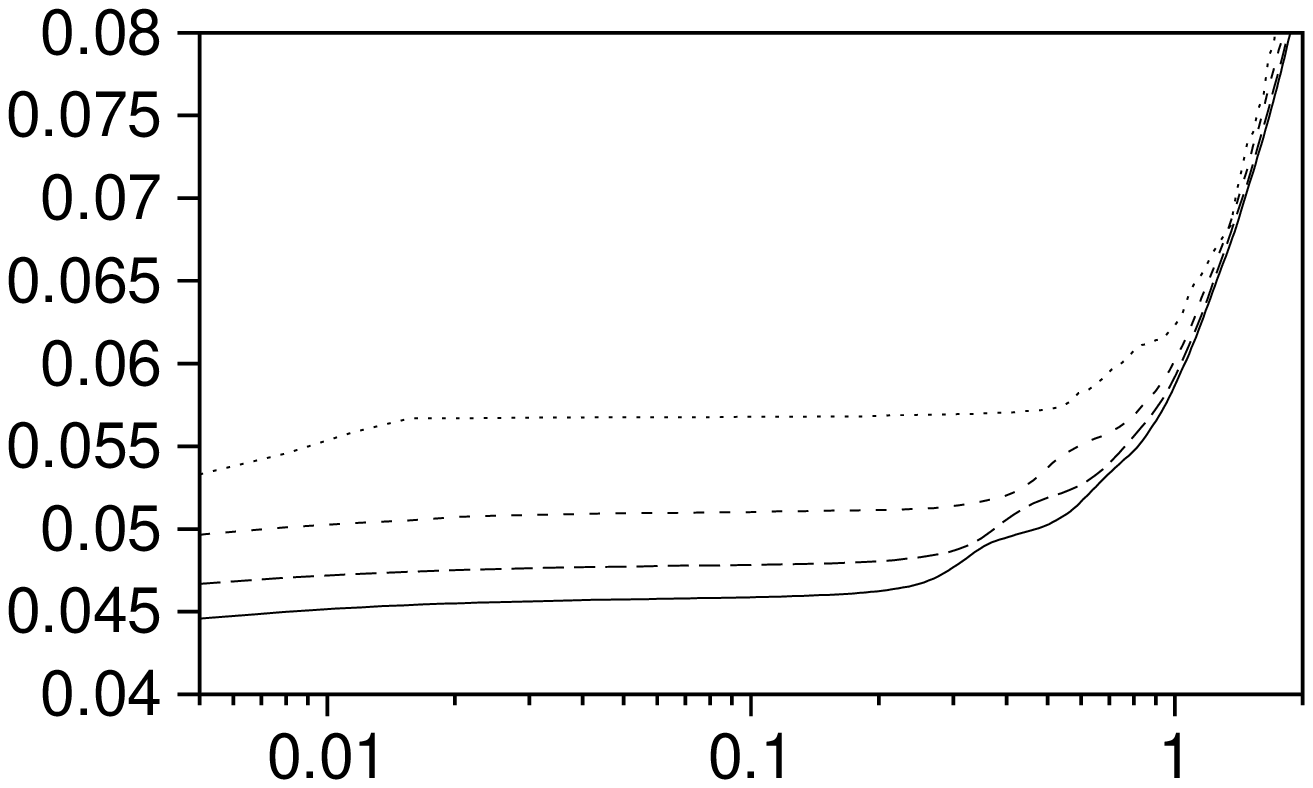}}
\vspace{0.4in}
}}}
\caption{We show the integrated conductance $I^L(\omega)$ for a 
quarter filled $\rho=1/4$ cyclic chain at $t=1,V=2$ (in non-ergodic regime) 
and for different sizes $L=24,20,16,12$. 
In the inset we show the same plot in semi-log scale in order to
illustrate the convergence of the charge stiffness. Observe that the
size of the horizontal plateau at small frequencies shrinks as
$\sim 2\pi/L$.}
\label{fig:9}
\end{figure}

An interesting conjecture has been put forward in Refs.\cite{Prelovsek}
(and critically debated in \cite{Kirchner,OnZotosConj}), namely that 
half-filled $\rho=1/2$ integrable 
t-V model should exhibit properties of an ideal insulator at all 
temperatures when $V>t$ (in our notation). Insulating behavior is 
characterized by $D^\infty_J = I^{\infty}(0) = 0$ and $(2/\beta)\sigma'(0) = 
(d/d\omega) I^{\infty}(0) = 0$, so the time correlation function 
$C_J(\tau)$ should be an oscillatory function in order to be integrated to zero.

We have found numerical evidence of (at least approximately) insulating behavior 
even in the non-integrable half-filled KtV model, when $V>t$.
In Fig.10 we demonstrate a double transition from (approximately) insulating regime
(example for $t=0.4,V=1$) to ideally conducting regime (example for $t=V=1$) to 
normally conducting regime (example for $t=4.5,V=1$) for a half-filled KtV model 
on $L=16$ sites.

\begin{figure}[htbp]
\hbox{\hspace{-0.1in}\vbox{
\hbox{
\leavevmode
\epsfxsize=3.6in
\epsfbox{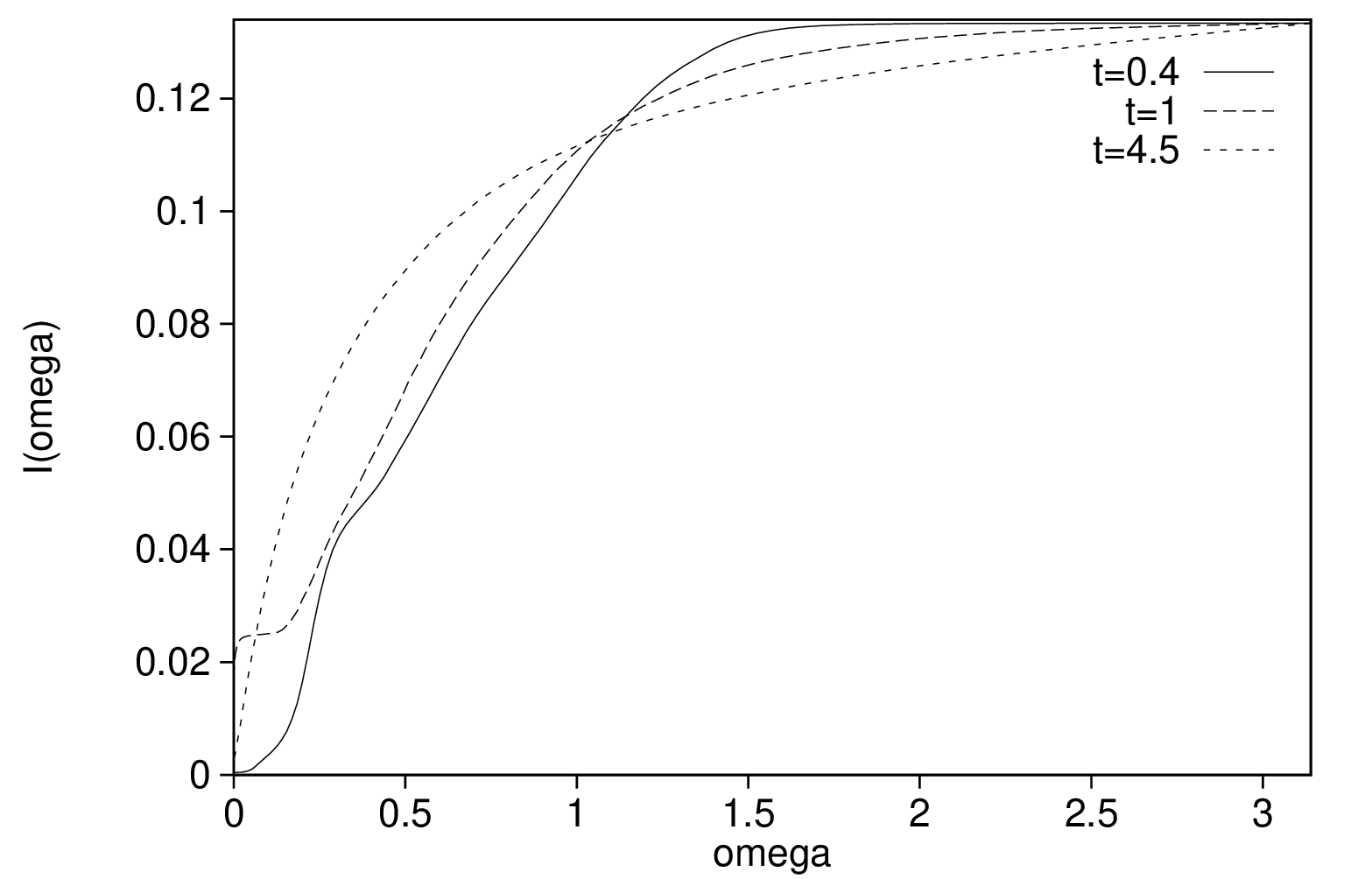}}
\vspace{-1.9in}
\hbox{\hspace{1.13in}
\vbox{
\leavevmode
\epsfxsize=2.25in
\epsfbox{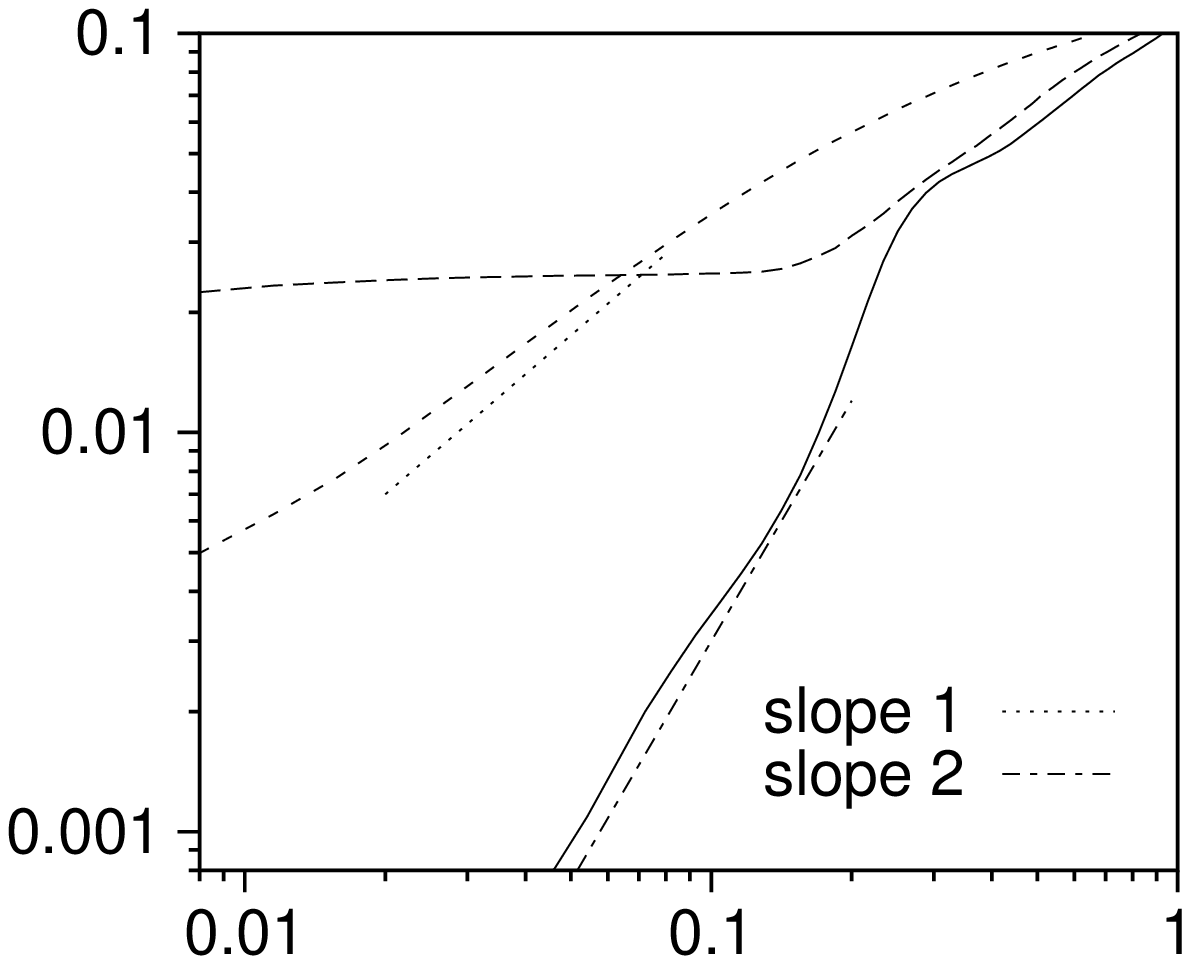}}
\vspace{0.45in}
}}}
\caption{Integrated conductance $I^L(\omega)$ for a 
half-filled ($\rho=1/2$) cyclic chain of size $L=16$
is shown for three different values of parameter $t$ (and fixed parameter $V=1$)
demonstrating a double transition from 
insulator (here for $t=0.4$) to ideal conductor (here for $t=1$) to a normal
conductor (here for $t=4.5$) as the kick parameter $t$ is increased.
In the inset we show the same three curves in log-log scale.}
\label{fig:10}
\end{figure}

\section{Third Method: Algebraic construction of time-averaged
observables for infinite system}

A large amount of numerical evidence has been presented
in previous two sections in support of the Conjecture put forward in Sec.I. 
However, all this evidence is based on computations on many-body systems 
of finite size $L$, and TL has been speculated by extrapolation to $1/L=0$. 
One may still have doubts, that in a non-integrable system that is close to
an integrable one quantum ergodicity may squeeze in for large sizes $L$, beyond
the scope of numerical observation. Therefore, as a complementary alternative,
one would like to have a method of computation of time-correlators, like 
$D_A$ (\ref{eq:stifA}), which would directly operate with infinite systems
on infinite lattices $L=\infty$. In this section we elaborate such a
method of computation of operator valued time-average of an observable 
$A$ in Heisenberg representation
\begin{equation}
\bar{A} = \lim_{M\rightarrow\infty} \frac{1}{2M+1}\sum_{m=-M}^M A(m).
\label{eq:ta}
\end{equation}
The method is specially designed for kicked systems whoose propagators
can be decomposed in several non-commuting parts \cite{TimeAverage}, and will
be implemented to compute time-averaged observables in infinite KtV model,
in particular $\bar{J}$ and $\bar{T}'$, and
the corresponding correlators, such as the charge stiffness $D_J$. 

\subsection{Mathematical structures}

The first essential mathematical structure used in this section is the
Hilbert space of {\em pseudo-local} quantum observables. Even in the
general setting we assume that the evolution propagator preserves the
number of particles, $[U,\rho]=0$.
So we again fix the density of particles $\rho$ and consider observables 
$A$ over a Fock (sub)space of quantum states with a given density 
parameter $\rho$. Such observables preserve the number of 
particles, $[A,\rho]=0$.

Let us define the {\em scalar product} of two {\em extensive} observables 
$A$ and $B$ as
\begin{equation}
(A|B) = \lim_{L\rightarrow\infty}\frac{1}{L}\ave{A^\dagger B}^L_\rho = 
\ave{\oneL A^\dagger B}_\rho
\label{eq:scp}
\end{equation}
We note that (\ref{eq:scp}) has all the necessary properties of a 
scalar (inner) product: it is linear in the right factor, positive,
and $(A|B)=(B|A)^*$. Note also that averaging over half-filled states is
in TL equivalent to the `grand-canonical' average, 
$\ave{.}_{\rho=\half}\equiv\ave{.}$.

The observable $A$ is called weakly local or pseudo-local, if
$\Vert A\Vert^2 := (A|A) < \infty$. Pseudo-local observables $A$ 
constitute a Hilbert space
denoted by ${\frak U}$. There is a {\em linear subspace} 
${\frak U}' \subseteq {\frak U}$ of pseudo-local observables $A$, 
such that $[A,B] = AB - BA$ is pseudo-local for any pseudo-local 
$B\in {\frak U}$. For any such $A\in {\frak U}'$, the scalar 
product (\ref{eq:scp}) is an invariant bilinear form with 
respect to the {\em adjoint map} $(\ad A)B = [A,B]$, namely
\begin{equation}
((\ad A^\dagger)B|C) = (B|(\ad A)C).
\end{equation}

The second essential mathematical structure is the {\em unitary}
Heisenberg-Floquet map $\hat{U}_{\ad} : {\frak U}\rightarrow{\frak U}$,
which propagates quantum observables in Heisenberg representation for
one period of time, starting at some time $\eta\in[0,1)$,
\begin{eqnarray}
\hat{U}_{\ad} A(\eta+m) = A(\eta+m+1) &=& U^\dagger A(\eta+m) U, \quad\\
(\hat{U}_{\ad} A|\hat{U}_{\ad} B) &=& (A|B). \label{eq:unitarity}
\end{eqnarray}
For example, for the KtV model (in spin representation which is,
for algebraic convenience,
used in this section) we have (\ref{eq:XXZ})
\begin{equation}
U\vert_{\eta=\half}=
\exp(-i\half t H_1)\exp(-i V (H_0 + \half S_z))\exp(-i\half t H_1)
\label{eq:evol2}
\end{equation}
with the potential and kinetic generator
\begin{eqnarray}
H_0 &=& \quar\sum_{j=-\infty}^\infty \sigma^z_j \sigma^z_{j+1},
\label{eq:gen}\\
H_1 &=& \quar\sum_{j=-\infty}^\infty
(\sigma^+_j \sigma^-_{j+1} + \sigma^-_j \sigma^+_{j+1}). \nonumber
\end{eqnarray}
Unlike in Sec.III, we have here taken the time steps
in the middle between the kicks, $\eta=\half$, in order to fully exploit
the time-reversal symmetry of the problem.
Note that the time evolution of observables which are diagonal in
momentum representation, like $J$ and $T'$, is not affected
by the shift $\eta$ of the origin of the stroboscopic map.
The Floquet-Heisenberg map can be written explicitly using an exponential
of the adjoint map
\begin{equation}
\hat{U}_{\ad} =
\exp(i\half t \ad H_0)\exp(i V \ad H_1)\exp(i\half t\ad H_0).
\label{eq:Uad}
\end{equation}
Since the density $\rho$ (or magnetization 
${\cal M}=\rho-\half$ in spin-$\half$ 
formulation) is fixed, the total spin $S_z$ in eq. (\ref{eq:evol2})
generates an irrelevant phase which does not influence the evolution of 
observables.

Time-average of observable (self-adjoint operator) $A$ (\ref{eq:ta}) is a
solution of the fixed-point equation for the Floquet-Heisenberg map
\begin{equation}
\hat{U}_{\rm ad} \bar{A} = \bar{A}.
\label{eq:invar}
\end{equation}
Time-averaging in operator space can also be written in terms of
an orthogonal projector $\hat{P}_U$ onto the null space
of $1-\hat{U}_{\rm ad}$, namely
\begin{equation}
\bar{A} = \hat{P}_U A,\quad
\hat{P}_U = \lim_{M\rightarrow\infty}\frac{1}{2M+1}\sum_{m=-M}^M
\hat{U}^m_{\rm ad}.
\label{eq:tal}
\end{equation}
The property
\begin{equation}
\hat{P}_U = \hat{P}_U^2
\end{equation}
is easily proved by writing a time-average limit (\ref{eq:tal}) 
in an equivalent, Gaussian way
$$\hat{P}_U = \lim_{M\rightarrow\infty}\frac{1}{\sqrt{2\pi}M}
\sum_{m=-\infty}^\infty \exp(-\half (m/M)^2) \hat{U}^m_{\ad}.$$

Without loss of generality we will in the following assume that
observable $A$ is traceless $\ave{A}=0$, such as $J$ and $T'$.
Note that the generalized stiffness (\ref{eq:stifA}) can be written
simply as
\begin{equation}
D_A = (A|\bar{A}).
\label{eq:stifH}
\end{equation}

\subsection{Time-average from Variational Principle in operator space}

Scaled (or normalized) time-average of a self-adjoint operator $A=A^\dagger$,
$X=\bar{A}/\Vert\bar{A}\Vert$, can be obtained from a {\em variational
principle in operator space}, namely as an extremum ({\em maximum}) 
of an action $s(X)$
\begin{eqnarray}
\frac{\delta}{\delta X} s(X) &=& 0,\nonumber \\
s(X) &=& \half (X|A)(A|X) = \half |(X|A)|^2
\label{eq:action}
\end{eqnarray}
with constraints
\begin{eqnarray}
\Vert (1 - \hat{U}_{\rm ad}) X \Vert^2 =
(X|(1-\hat{U}^{-1}_{\ad})(1-\hat{U}_{\ad})|X) = 0, \label{eq:constr1} \\
(X|X) = {\rm const} < \infty,\label{eq:constr2}
\end{eqnarray}
Namely, eqs. (\ref{eq:action}-\ref{eq:constr2})
imply $X=\alpha \bar{A}$, where $\alpha = \Vert\bar{A}\Vert^{-1}$ if $X$ is
normalized $(X|X)=1$. 
Since constraint (\ref{eq:constr1}) is homogeneous the corresponding 
Lagrange multiplier is diverging. Hence, we suggest to write the
constrained variational problem (\ref{eq:action},\ref{eq:constr1},
\ref{eq:constr2}) in the compact form
\begin{eqnarray}
&& \lim_{\epsilon\rightarrow\infty}
\frac{\delta}{\delta X} s_{\epsilon}(X) = 0
\quad \Rightarrow \quad
X = \alpha \bar{A},\;\; \alpha\in\C \nonumber \\
&& s_\epsilon(X) = \half |(X|A)|^2 -
\half\lambda\Vert\epsilon^{-1}(1 - e^{-\epsilon}\hat{U}_{\ad})X\Vert^2
\label{eq:action2}
\end{eqnarray}
where $\lambda$ is another Lagrange multiplier associated with 
the second constraint (\ref{eq:constr2}).
Indeed, for small $\epsilon$ one may write
\begin{eqnarray*}
s_\epsilon(X) &=& \half |(X|A)|^2 - \half \lambda (A|A) \\
&-& \half\lambda (\epsilon^{-2}-\epsilon^{-1}) 
\Vert (1-\hat{U}_{\rm ad})X \Vert^2 + {\cal O}(\epsilon),
\end{eqnarray*}
so homogeneous constraint (\ref{eq:constr1}) follows automatically as 
$\epsilon\rightarrow 0$ in order to make the action $s_\epsilon(X)$ regular 
(and maximal) at $\epsilon=0$.
Let us now show that the above variational problem has the
correct solution (\ref{eq:ta}). We differentiate the
action (\ref{eq:action2})
\begin{eqnarray*}
\frac{\delta}{\delta X} s_\epsilon(X_\epsilon) &=&
(A|X_\epsilon) A - \frac{\lambda}{\epsilon^2}
(1-e^{-\epsilon}\hat{U}^{-1}_{\ad})
(1-e^{-\epsilon}\hat{U}_{\ad})X_\epsilon\\ 
&=& 0
\end{eqnarray*}
and write $a=(A|X)$.
This equation can be solved explicitly for $X_\epsilon$
\begin{eqnarray}
X_\epsilon &=& \frac{a\epsilon^2}{\lambda}
(1-e^{-\epsilon}\hat{U}_{\ad})^{-1}
(1-e^{-\epsilon}\hat{U}^{-1}_{\ad})^{-1} A \nonumber\\
&=& \frac{a\epsilon^2}{\lambda}
\sum_{n=0}^\infty \sum_{m=0}^\infty e^{-(n+m)\epsilon} 
\hat{U}^{n-m}_{\ad}A \nonumber\\
&=& \frac{a\epsilon^2}{\lambda}
\sum_{p=-\infty}^\infty A(p) \sum_{q=|p/2|}^\infty e^{-q\epsilon} 
\nonumber \\
&=& \frac{a\epsilon^2}{\lambda (1-e^{-\epsilon})}
\sum_{p=-\infty}^\infty A(p) e^{-(\epsilon/2)|p|}.
\label{eq:lastexpr}
\end{eqnarray}
In the limit $\epsilon\rightarrow 0$, the last expression 
(\ref{eq:lastexpr}) is proportional to the time-average
\begin{equation}
X = \lim_{\epsilon\rightarrow 0} X_\epsilon =
\frac{4 a}{\lambda}\bar{A}
\end{equation}
Evaluation
$$ a =(A|X) = a \frac{4 D_A}{\lambda} $$
fixes the value of the Lagrange multiplier
\begin{equation}
\lambda = 4 D_A.
\end{equation} 
Unitarity (\ref{eq:unitarity}) and invariance (\ref{eq:invar})
have the following very important consequence
\begin{equation}
(\bar{A}|\bar{A}) = (A|\bar{A}) = (\bar{A}|A).
\end{equation}
Assuming that $X$ is non-vanishing so that $(X|X) > 0$,
and that $X$ and $\bar{A}$ are {\em proportional}, one can write
\begin{equation}
\bar{A} = \frac{(X|\bar{A})}{(X|X)} X = \frac{(X|A)}{(X|X)} X.
\end{equation}
Taking the scalar product of the last equation with $A$, one obtains
a very useful expression for the stiffness
\begin{equation}
D_A = \frac{|(A|X)|^2}{(X|X)}.
\label{eq:stifX}
\end{equation}

\subsection{Numerical application}

However, the maximization of the functionals (\ref{eq:action},
\ref{eq:action2}) in the huge infinitely dimensional operator
space ${\frak U}$ is not convenient for practical calculation.
Instead, we suggest to estimate the time-averaged observable
$\bar{A}$ by solving the variational problem (\ref{eq:action2})
in a finite-dimensional subspace ${\frak M}(A) \subset {\frak U}$.
(Galerkin-like approach in operator space).
In fact, we devise a special sequence of truncated
`minimal invariant' operator spaces $\ldots {\frak M}_p(A) \subset
{\frak M}_{p+1}(A) \ldots \subset {\frak U}$, which in the limit
$p\rightarrow\infty$ (after closure) contain the time average $\bar{A}$.
Thus, the solutions $X_p$ of the variational problems
(\ref{eq:action2}) on spaces ${\frak M}_p$ should converge
to the proper scaled time-average $X$ of observable $A$.

Let ${\frak s} = \{\alpha H_0 + \beta H_1; \alpha,\beta\in\C\}$
be 2-dim linear vector space spanned by the two generators of motion, 
$H_\alpha,\; \alpha=0,1$ (\ref{eq:gen}). 
Let us define the {\em minimal invariant
operator space containing} $A$, as the closure of linear combinations of all
products of adjoint generators $\ad H_\alpha,\;\alpha=0,1$ on $A$
\begin{equation}
{\frak M}(A) = \overline{\bigcup_{n=0}^\infty (\ad{\frak s})^n A}.
\end{equation}
${\frak M}(A)$ is indeed the minimal (though infinitely-dimensional
in general) operator space containing $A$ with the invariance property
\begin{eqnarray}
(\ad H_\alpha) {\frak M}(A) &=& {\frak M}(A),\quad\alpha=0,1, \nonumber \\
\hat{U}_\ad {\frak M}(A) &=& {\frak M}(A),\quad\forall t,V.
\end{eqnarray}
It is obvious that $\bar{A}\in {\frak M}(A)$.
We now construct the countable basis of the space ${\frak M}(A)$ ordered
by the order of locality as follows: We assign an observable
$\tilde{H}_{q,c}$ to an {\em ordered pair} of integers $(q,c)$,
{\em order} $q$, and {\em code} $c,\, 0\le c < 2^{q-1}$ with $q-1$
binary digits $c_n,\,c=\sum_{n=1}^{q-1} c_n 2^{n-1}$, namely
\begin{equation}
\tilde{H}_{q,c} =
(\ad H_{c_{q-1}})(\ad H_{c_{q-2}})\cdots(\ad H_{c_1}) A.
\label{eq:tilH}
\end{equation}
Since not all observables $\tilde{H}_{q,c}$ upto a given maximal
order $p, q\le p,$ are linearly independent we perform Gram-Schmit
orthogonalization w.r.t. the scalar product (\ref{eq:scp})
\begin{eqnarray}
G_{q,c} &=& \cases{
\tilde{G}_{q,c}/\sqrt{(\tilde{G}_{q,c} | \tilde{G}_{q,c})};
& $\tilde{G}_{q,c} \neq 0$,\cr
0; & $\tilde{G}_{q,c} = 0$,\cr }
\nonumber \\
\tilde{G}_{q,c} &=& \tilde{H}_{q,c} -
\sum_{(r,b)}^{(r,b)<(q,c)}
G_{r,b} (G_{r,b}|\tilde{H}_{q,c}).
\label{eq:base}
\end{eqnarray}
The nonzero observables
$G_{q,c}$ form the orthonormal basis of ${\frak M}(A)$.
Note that observables $G_{q,c}$ are strictly local
operators of order $q$: in case of spin representation of KtV model
they are represented as expansions
\begin{equation}
G_{q,c}=\sum\limits_{s_0,s_1,\ldots s_q} g_{q,c}^{s_0 s_1 \ldots s_q}
Z_{s_0 s_1 \ldots s_q} $$
\label{eq:expan}
\end{equation}
in terms of spatially homogeneous finite products of
field operators
$$ Z_{s_0 s_1\ldots s_q}=\sum\limits_{j=-\infty}^\infty
\sigma_j^{s_0}\sigma_{j+1}^{s_1}\cdots
\sigma_{j+q}^{s_q},$$
where $s_k\in\{0,+,-,z\}$ and $\sigma_j^0=1$.
The (average) number of nonzero terms in expansions (\ref{eq:expan})
was found to grow exponentially at approximately the
same rate for both observables under study, $J$ and $T'$, namely as
$$ \#\{ g_{q,c}^{s_0 s_1 \ldots s_q} \neq 0\} \approx 0.5\,\times\,2.55^q,$$
which may be further reduced by a factor 2, or even by a factor 4 if
$\rho=\half$, using the symmetries $\hat{\cal P}$ (\ref{eq:parity}),
and $\hat{\cal R}$ (\ref{eq:parthole}) 
(the latter may be used if $\rho=\half$). 
Note that the entire linear space ${\frak M}(A)$
has the same symmetry properties as observable $A$, for example
the space ${\frak M}(J)$ and ${\frak M}(T')$ belong to a negative and
positive parity symmetry class, respectively, 
w.r.t. parity operation (\ref{eq:parity}),

Let us now define a sequence of {\em truncated minimal invariant
operator spaces} containing $A$,
\begin{equation}
{\frak M}_p(A) = \bigcup_{n=0}^{p-1} (\ad{\frak s})^n A,\quad p=1,2\ldots
\end{equation}
with dimensions $d_p(A) := \dim{\frak M}_p(A)$.
Linear space ${\frak M}_p(A)$ contains operators derived
from $A$ by composition of generators $\ad H_\alpha$ up to {\em order} $p$.
Due to binary code construction (\ref{eq:tilH})
we have a strict upper bound on the growth of dimensions of the
truncated spaces ${\frak M}_p(A)$,
\begin{equation}
d_p(A) \le 2^{p-1},
\label{eq:dpa}
\end{equation}
however actual dimensions may grow considerably slower (due to
systematic linear dependences among $\tilde{H}_{p,b}$), namely for
$A=J$ and $A=T'$ we find by computer algebra up to $p=14$th order (see Tab.1)
\begin{equation}
d_p(J) \approx 1.825^{p-1},\quad
d_p(T') \approx 1.68^{p-1}.
\label{eq:dpa2}
\end{equation}
Let $\ma{H}_{p,\alpha}$, $\alpha=0,1$, denote real and symmetric
(Hermitean in general)
matrices of linar maps $\ad H_\alpha$ on ${\frak M}_p(A)$ with
images orthogonally projected back to ${\frak M}_p(A)$.
It follows from the construction that they have (generally)
a block-banded structure where the blocks correspond to
observables with a fixed order $q$: namely
\begin{equation}
(G_{q,c}|\ad H_\alpha|G_{q',c'}) = 0,\quad {\rm if}\quad
|q-q'|\neq 1.
\end{equation}
The truncated adjoint maps, $\ma{H}_{p,\alpha}$,
have nontrivial null spaces
\begin{equation}
{\frak N}_{p,\alpha}(A) =
\{ B\in {\frak M}_p(A); [H_\alpha,B]\in
{\frak M}_{p+1}(A)-{\frak M}_p(A)\},
\end{equation}
with dimensions $d_{p,\alpha}(A):=\dim{\frak N}_{p,\alpha}(A)$ which
increase approximately with the same exponent as $d_p(A)$
(\ref{eq:dpa2}) (see Tab.1). 

By means of truncated adjoint maps $\ma{H}_{p,\alpha}$ we
construct an approximate Floquet-Heiseberg matrix $\ma{U}_p$, which
is a $d_p(A)$-dim {\em unitary} 
matrix over the truncated space ${\frak M}_p(A)$
\begin{equation}
\ma{U}_p = \exp(i\half t \ma{H}_{p,1})
           \exp(iV \ma{H}_{p,0})
           \exp(i\half t \ma{H}_{p,1})
\end{equation}
Now we are ready to solve the variational problem
(\ref{eq:action}-\ref{eq:action2}) in the truncated space ${\frak M}_p(A)$.
We note an important `experimental' observation (whose theoretical
understanding is stil lacking) namely that the matrix
$\ma{1}-\ma{U}_p$ possesses a high-dimensional null-space
$${\frak N}^U_p(A) = \{B\in {\frak M}_p(A); \ma{U}_p B = B\},$$
whose dimension $d^U_p(A) := \dim{\frak N}^U_p(A)$ is, for odd $p$, 
independent of parameters $t,V$ and equal to the dimension of the 
null space of $\ma{H}_{p,1}$,
\begin{equation}
d^U_{2l-1}(A)= d_{2l-1,1}(A).
\end{equation}
Note also that for odd order of truncation $p=2l-1$, the elements of
null space $B\in {\frak N}^U_p(A)$ are spanned by combinations
of {\em even} powers of generators only, i.e.
$(B|G_{2l,c})\equiv 0$, which is due to
time-symmetric construction ($\eta=\half$) of the evolution operator 
$\hat{U}_{\ad}$ (\ref{eq:Uad}).

The scalar products $(\ref{eq:scp})$ for different values of the density 
$\rho$ are non-degenerate with respect to each other, and therefore 
the dimensions of various linear (sub)spaces, 
$d_p(A),d_{p,\alpha}(A),d^U_p(A)$, (see Tab.1)
{\em do not depend} on the density parameter $\rho$.

The constraint (\ref{eq:constr1}) is now equivalent to
restricting the variation (\ref{eq:action}) 
to the sub-space ${\frak N}^U_p(A)$.
Hence, the `truncated' scaled time-averaged observable $X_p$ is
a maximum of the quadratic form $(X_p|A)(A|X_p)$ on ${\frak N}^U(A)$ with
a normalization constraint $(X_p|X_p)=1$. In other words, if
$F_n,n=1,\ldots,d:=d^U_p(A)$ is an orthonormal basis
of the null-space ${\frak N}^U_p(A)$ and if $(x_1,\ldots,x_d)$
is a normalized eigenvector of the (positive definite) $d\times d$ matrix
eigenvalue problem
$$\sum_n (F_m|A)(A|F_n) x_n = f x_m$$
with the {\em maximal} eigenvalue $f$, then
\begin{equation}
X_p = \sum_n F_n x_n
\end{equation}
is a solution of the variational problem (\ref{eq:action}-\ref{eq:action2})
in the truncated space ${\frak M}_p(A)$. In the limit $p\rightarrow\infty$
we expect to recover an exact time-average
\begin{equation}
\lim_{p\rightarrow\infty} X_p = X = \Vert A\Vert^{-1} \bar{A}.
\label{eq:limit}
\end{equation}
However, if the system is ergodic, time average should be zero 
$\bar{A}=0$ (note that $\ave{A}=0$), so the (normalizable) limit 
(\ref{eq:limit}) should not exist.
In order to inspect the convergence of $X_p$ in the Hilbert space of 
observables ${\frak U}$, we
define a {\em relative norm} $N_q(X)$ w.r.t. order $q$
\begin{equation}
N_q(X) = \sum_c |(X|G_{q,c})|^2. 
\end{equation}
Since 
\begin{equation}
\Vert X\Vert^2 = \sum_{q=0}^\infty N_q(X),
\label{eq:ssum}
\end{equation}
the inspection of the convergence of the sum on RHS of (\ref{eq:ssum}) would
give us indication of convergence of $X_p$ (\ref{eq:limit}) and thus 
non-ergodicity of the problem.
As a second criterion of convergence of $X$ we study {\em stability} of
$X_p$, or of the relative norms $N_q(X_p),q \le p$, w.r.t. variation of
the truncation order $p$.

\begin{table}
\begin{tabular}{||r|r|r|r|r|r|r||} 
$p$ & $d_p(J)$ & $d_{p,0}(J)$ & $d_{p,1}(J)$ & 
    $d_p(T')$ & $d_{p,0}(T')$ & $d_{p,1}(T')$ \\ \hline  
1  & 1    & 1   &  1   & 1   & 0   & 1  \\
2  & 2    & 0   &  2   & 2   & 0   & 1  \\
3  & 4    & 2   &  2   & 4   & 1   & 2  \\
4  & 7    & 3   &  3   & 6   & 2   & 1  \\
5  & 12   & 6   &  4   & 10  & 4   & 5  \\
6  & 21   & 9   &  7   & 15  & 5   & 2  \\
7  & 38   & 16  &  12  & 25  & 9   & 10 \\
8  & 69   & 27  &  21  & 40  & 12  & 7  \\
9  & 126  & 48  &  38  & 66  & 22  & 21 \\
10 & 230  & 84  &  68  & 107 & 33  & 22 \\
11 & 419  & 153 &  123 & 178 & 60  & 51 \\
12 & 763  & 273 &  223 & 293 & 91  & 66 \\
13 & 1393 & 493 &  409 & 494 & 162 & 137\\
14 & -    & -   &  -   & 831 & 257 & 202\\
\end{tabular}
\caption{Dimensions of the truncated minimal invariant spaces and
of the null-spaces of truncated adjoint maps for different orders
of truncation $p$.}
\label{tab:1}
\end{table}

In Figs.11,12 we show the relative norms $N_q(X_p)$ of
the normalized time-average of the current $J$ (Fig.11) and kinetic energy $T'$ (Fig.12)
for several different orders of truncation $p$ (up to $p=11$ for $J$ and
up to $p=13$ for $T'$). We note that for both observables, $X_p$ is
quite stable against variation of $p$, for $t \le 1.4$, and in the
same parameter range, the coefficients $N_q(X_p)$ seem to be summable.
The stiffness $D_A$ (\ref{eq:stifX}) may be re-written in terms of 
relative norms as
\begin{equation}
D_A = (A|A) \frac{N_1(X)}{\sum_{q=1}^\infty N_q(X)}.
\label{eq:da}
\end{equation}
In this regime where $X$ is convergent in $\frak{U}$, $N_q(X_p)$ are good 
approximation of $N_q(X)$ for $q \le p$ (appart from a constant renormalization 
prefactor which cancels out from expressions like (\ref{eq:da})), 
and we may write a good estimate for the upper bound on the stiffness
\begin{eqnarray}
D^p_A &=& |(A|X_p)|^2 \approx (A|A) \frac{N_1(X)}{\sum_{q=1}^p N_q(X)} \nonumber\\
      &>& (A|A) \frac{N_1(X)}{\sum_{q=1}^\infty N_q(X)} = D^\infty_A.
\end{eqnarray}
However, we would like to have accurate approximations of the stiffness 
$D^{\infty}_A$ itself rather than the accurate upper bounds, so we extrapolate the
relative norms $N_q(X)\approx N_q(X_p)$ to orders higher than the
order of truncation, $q > p$, in expression for the stiffness (\ref{eq:da})
by fitting the tail of $N_{q=2l+1}(X_p)$ at three points: 
$q=p-4,p-2,p$ (note that $N_{q=2l}(X_p)=0$) 
with exponential ansatz $N_{q=2l+1}(X_p)\propto\exp(-s q)$.
Since the actual rate of convergence of $N_q(X)\rightarrow 0$, as 
$q\rightarrow\infty$ is probably slower than exponential (see Figs.11,12), 
the stiffness extrapolated in this way (eq. (\ref{eq:da})), 
$D^e_A$, is probably still slightly overestimated.
In Fig.13 we show the dependence of the extrapolated charge stiffness $D^e_J$
on the parameter $t$ (and for fixed $V=2$) through the critical range
$t \sim t_c \approx 1.45$, and compare it with the charge stiffness
as computed from direct diagonalization of the finite KtV chains of sizes 
$L=24,20,16$. When approaching the critical point $t_c$, the fitted 
slope $s$ linearly decreases to zero.
For larger values of parameters, $t > t_c(V)$,
$X_p$ is not stable against variation of $p$ and the partial sums of 
relative norms $N_q(X)$ are not converging. 
Therefore, $\bar{A}=0$ and $D^\infty_A=0$, and the system 
is quantum ergodic. In Fig.14 we plot a full $(t,V)$ phase diagram 
of the (extrapolated) charge stiffness $D^e_J$. 
It is clear that this last method, since in operates with 
an infinite system, gives the most reliable results on the
critical regions of transition between dynamical phases.
However, no other dynamical information on correlation functions is obtained
other than their time-averages, so within the present method we cannot make any
claims on the stronger ergodic property of quantum mixing.

Although dynamical behavior of observables may depend on the symmetry class
of observables w.r.t., say, parity operation (\ref{eq:parity}), we have found very 
similar ergodic, non-ergodic, and critical regions, for the two examples of opposite 
parity observables, $J$ and $T'$, that have been studied.
However, we should note that dynamical behavior may also depend on the other
(continuous) conserved quantities, such as the density $\rho$. Our results for other
values of density $\rho$ indicate that the transition region between ergodic and 
non-ergodic dynamics moves to slightly smaller values of parameter $t$ as $\rho$ 
approaches $1/2$. Due to paricle-hole transformation (\ref{eq:parthole}) the 
dynamics for $\rho=\rho'$ is equivalent to dynamics for $\rho=1-\rho'$.

\begin{figure}[htbp]
\begin{center}
{\leavevmode
\epsfxsize=3.2truein
\epsfbox{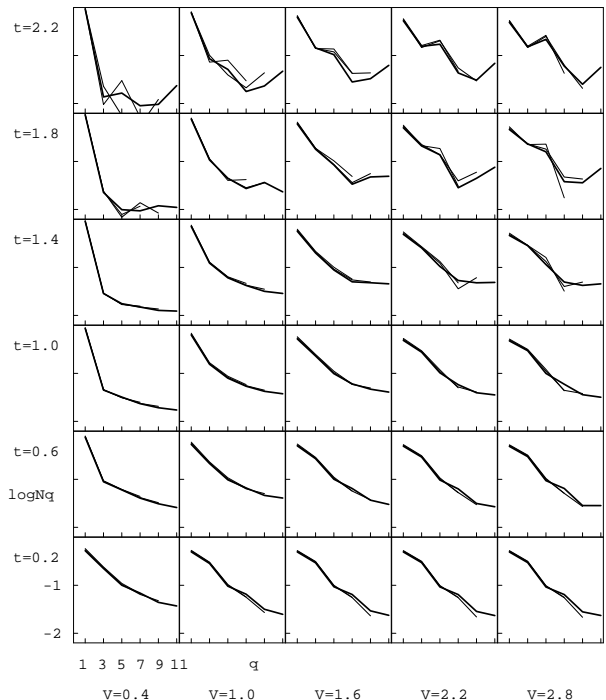}}
\end{center}
\caption{The logarithm of relative norm $N_q(X)$ of the
normalized time-averaged current $X=\bar{J}/\Vert J\Vert$
in a quarter-filled ($\rho=1/4$) infinite KtV model is plotted against 
(odd) order $q=2l-1$, for a square mesh of parameters $t$ and $V$. 
The three different curves on each graph, thick, medium, and thin, 
refer to three different orders of truncation of operator spaces, 
$p=11$, $p=9$, and $p=7$, respectively.}
\label{fig:11}
\end{figure}

We should note that in a recent paper \cite{Prosenktvi} a
very similar algebraic approach has been used in order to compute the
pseudo-local quantum invariants of motion. 
In the regime of non-ergodic dynamics one or two converged pseudo-local 
invariants of motion were found, whereas in the regime of ergodic dynamics, 
no non-trivial invariants of motion were found.
Then by using a formula (\ref{eq:Mazur}) of Mazur \cite{Mazur} 
and Suzuki\cite{Suzuki}, the time averaged correlation of kinetic energy
$D_T$ has been computed by means of an expansion in terms of pseudo-local
invariants, giving the results which are in good agreement with the results of 
direct calculations on finite systems.
We believe that our variational approach in the space of observables presented
here is (in general, possibly non-integrable case) an improvement of the Mazur-Suzuki
approach \cite{Mazur,Suzuki} to the calculation of time-averaged correlators.
Within Mazur formula (\ref{eq:Mazur}) one is typically able to write only 
inequality (lower bound on stiffness) since 
the set of {\em known local} invariants of motion may be {\em incomplete}.

\begin{figure}[htbp]
\begin{center}
{\leavevmode
\epsfxsize=3.2truein
\epsfbox{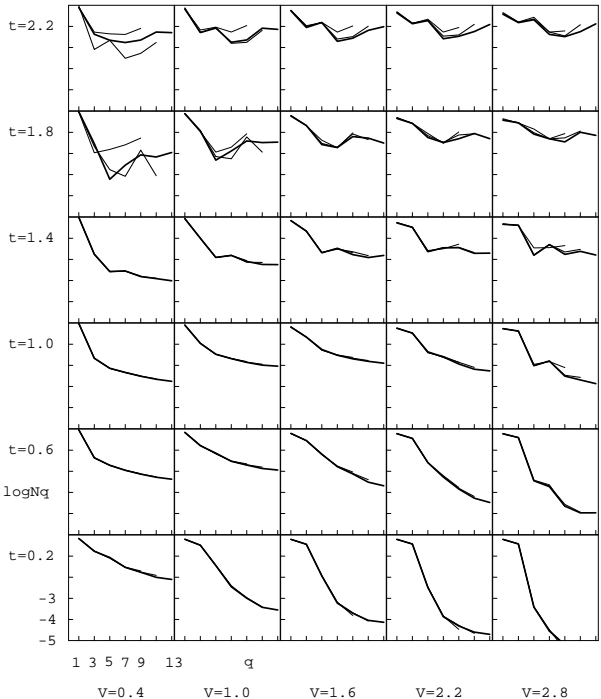}}
\end{center}
\caption{Same as in previous Fig.11 for the
normalized time-averaged kinetic energy  $X=\bar{T}'/\Vert T'\Vert$,
except for larger truncation orders, $p=13$ (thick curves), $p=11$ (medium curves),
$p=7$ (thin curves).}
\label{fig:12}
\end{figure}

\begin{figure}[htbp]
\hbox{\hspace{-0.1in}\vbox{
\hbox{
\leavevmode
\epsfxsize=3.6in
\epsfbox{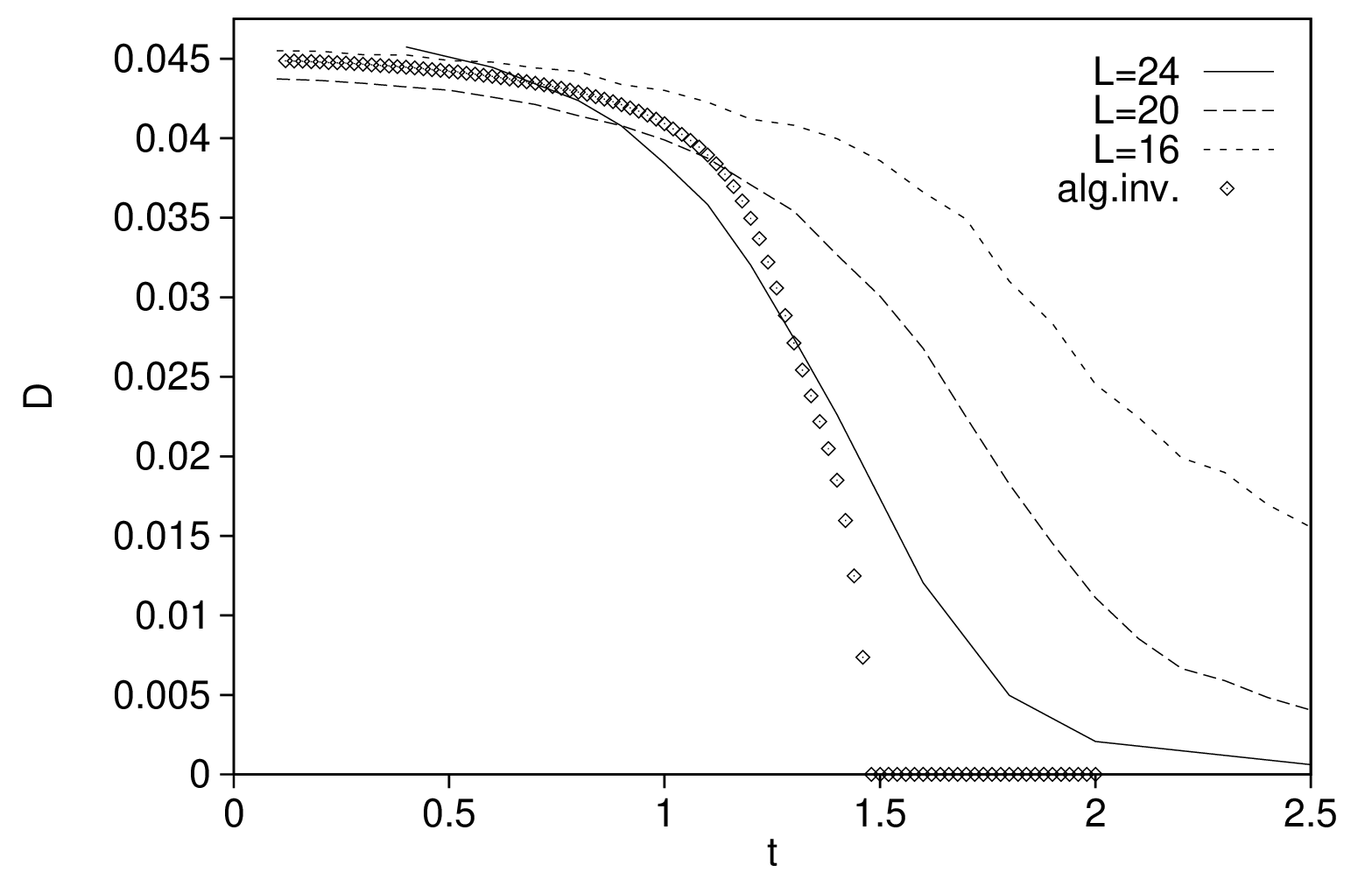}}
\vspace{-1.43in}
\hbox{\hspace{0.25in}
\vbox{
\leavevmode
\epsfxsize=1.6in
\epsfbox{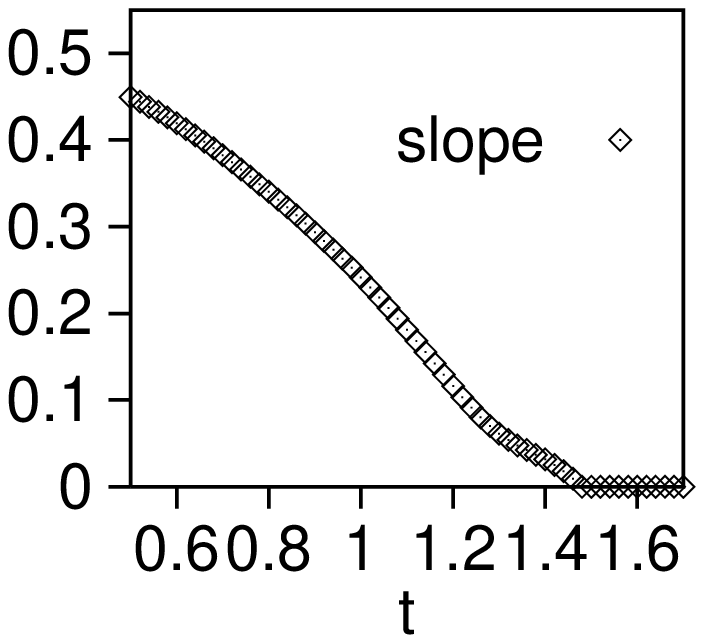}}
\vspace{0.4in}
}}}
\caption{Charge stiffness $D_J$ vs. kick parameter $t$ and
constant parameter $V=2$ for quarter-filled chain $\rho=1/4$.
Different curves refer to different system sizes
$L=24,20,16$, while points represent the infinite-size stiffness 
(\ref{eq:stifX}) based on extrapolated algebraic time-averaged current
invariant of motion. The truncation order is $p=11$.
In the inset we show the logarithmic slope $s$ of the falloff of the relative 
norms $N_q(\bar{J})\propto \exp(-sq)$ at $q\approx p$.}
\label{fig:13}
\end{figure}

\begin{figure}[htbp]
\hbox{\hspace{-0.1in}\vbox{
\hbox{
\leavevmode
\epsfxsize=3.6in
\epsfbox{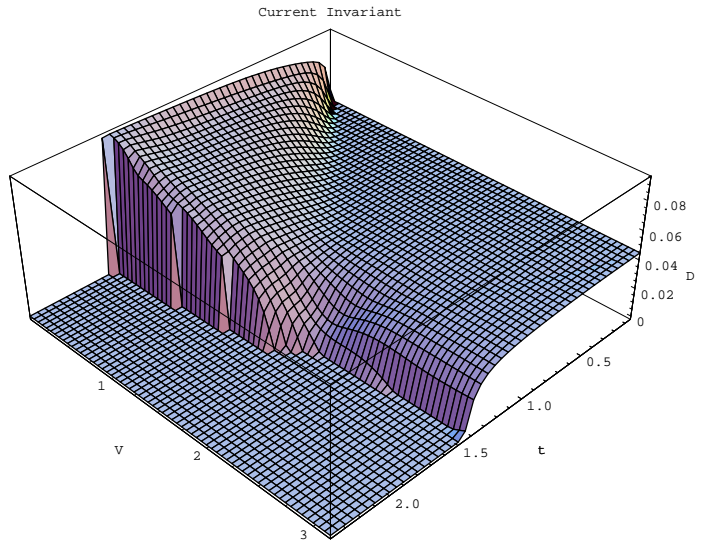}}
}}
\caption{$(t,V)$ phase diagram of the charge stiffness $D_J$ for quarter-filled
$(\rho=1/4)$ infinite KtV model, as deduced from {\em extrapolated}
time-averaged current $\bar{J}$.}
\label{fig:14}
\end{figure}

\section{Conclusions and Discussion}

In this paper we have presented three complementary (mainly numerical and
computer-algebraic) approaches to the dynamics of non-integrable quantum 
many-body systems in thermodynamic limit (TL), demonstrated and studied in a 
kicked t-V model of spinless fermions. We have been primarily interested in 
the structural stability of non-ergodic quantum motion and the transition 
from non-ergodic/non-mixing to ergodic/mixing dynamics in TL.

The first approach that we have used is a direct time-evolution of a finite 
quantum system (which may be in the present model performed very efficiently by 
means of the so-called Fermionic Fast Fourier Transformation) 
and computation of time-correlation
functions of generic quantum observables. The size $L$ of the system is 
systematically increased and TL is speculated based on extrapolation to $1/L=0$.
For sufficiently large values of kick parameters, we have
found quantum mixing and {\em exponential} decay of time correlation functions,
while for smaller, intermediate values ($\sim 1$) of kick parameters, we have
found non-mixing quantum motion characterized by saturating, non-vanishing 
time-correlation functions.

Our second approach is a direct diagonalization of the stationary quantum problem
of finite size and calculation of dynamical properties, such as charge stiffness,
conductivity, etc., in frequency domain. Also traditional quantum signatures of
chaos, such as level statistics, has been inspected and shown to correspond
with the dynamical behavior. This approach is less computationally efficient
in case of the present model than the first one.

In the third approach, which is fully complementary to the other two,
we propose an algebraic method for computation of time-averaged 
observables of an infinite 
system. Thus we can make most precise statements on quantum-ergodicity of 
an infinite system, which are in complete agreement with the extrapolated
results of calculations on finite systems.

The above results are claimed to be the evidence for the validity of the
Conjecture (Sec.I), namely that the generic quantum-many-body system in TL
may not be quantum-ergodic (or mixing) if it is sufficiently close to an
integrable system in parameter space. Recent numerical results on transport in 
extended (non-integrable) Hubbard model \cite{Kirchner} are compatible with the 
above Conjecture. The transition between non-ergodic and ergodic-dynamics when the 
external parameters are increased has the properties of (dynamical) phase
transition and should be further studied theoretically.
First such attempt has been undertaken in \cite{ProsenQFTmap}, where 
a discretized non-integrable quantum field model (in the continuum limit) has been 
mapped on a quantum chaotic model of a single particle on a 2-dim torus 
(in the quasi-classical limit),
and the transition from non-ergodic/non-mixing dynamics to
ergodic/mixing dynamics of the quantum field model has been
identified with the stochastic transition from regular to 
chaotic motion.

We have also given a clear evdience on the 
non-trivial existence of {\em mixing quantum 
motion} in KtV model in TL with {\em exponentially} decreasing time-correlation 
functions, provided the external (kick) parameters are large enough
(above the critical values).
Such quantum mixing behavior may be a source of truly {\em chaotic} and 
{\em macroscopically irreversible} quantum motion of many-body systems 
\cite{PUP}. Note that macroscopic irreversibility as a consequence of 
{\em non-dissipative} but strongly non-integrable quantum many-body 
dynamics has been recently observed experimentally\cite{Usaj}.  

One might doubtfully argue that our quite 
surprising finding on structurally
stable non-ergodic quantum motion in TL (formulated as Conjecture in Sec.I)
may be just another peculiarity of Physics in One-dimension, and as such should not
be expected to hold in interacting quantum systems in more than one spatial
dimension. Being aware of this fear we have straightforwardly extended our 
KtV model (\ref{eq:KtV1},\ref{eq:mbmap}) to a rectangular periodic $L_1\times L_2$ 
lattice in two spatial dimensions, with isotropic hopping in two orthogonal 
directions and $\delta-$kicked isotropic nearest neighbour interaction.
An efficient direct time evolution of the 2-dim KtV model has been
implemented analogously along the lines described in Sec.III, and 
its time correlation functions have been computed accordingly, however due to
greater computational complexity only for relatively small lattices of sizes up to
$6\times 5$. We should stress that we were again able to identify quite clearly 
the two regimes of quantum motion which have been roughly stable against the 
variation of the lattice size, namely: (i) a quantum mixing regime for 
sufficiently large $t$ 
and $V$, and more important, (ii) quantum non-ergodic and non-mixing regime 
for $|t|\sim |V|\sim \half$ (or smaller), although the system is not known 
to be analytically integrable in the limit $t\rightarrow 0,V\rightarrow 0$. 
This result (whose details will be published elsewhere) is a small
piece of numerical evidence for the validity of the Conjecture in 
two spatial dimensions.

Discussions with Prof. P. Prelov\v sek, and the financial support by the
Ministry of Science and Technology of R Slovenia are
gratefully acknowledged.


\begin{references}
\bibitem{Temp} We would like to demonstrate statistical laws with the
minimal external randomness. Hence we think of an isolated system
left to evolve with no contact with thermal reservoirs,
since we would like to inspect how non-linear interaction
alone can control the ergodic properties of dynamics.
\bibitem{TimeAverage} The simpler and perhaps more powerful version of
the method for the autonomous Hamiltonian systems, which may be
used also to give rigorous upper bounds on charge or spin stiffness etc., 
will be presented elsewhere.
\end{references}
\end{document}